\documentclass[%
reprint,
superscriptaddress,
amsmath,
amssymb,
aps,
floatfix,
showkeys
]{revtex4-2}

\usepackage{graphicx}
\usepackage{dcolumn}
\usepackage{bm}
\usepackage[dvipsnames]{xcolor}
\usepackage{float}
\usepackage{verbatim}
\usepackage{xspace}
\usepackage{siunitx}
\usepackage[version=4]{mhchem}
\usepackage{enumitem}
\usepackage{multirow}
\usepackage[
    colorlinks=true,
    linkcolor=NavyBlue,  
    citecolor=NavyBlue,  
    urlcolor=gray,       
]{hyperref}

\graphicspath{{./}{./figures}}

\newcommand{\model}[0]{CarNet\xspace}
\newcommand{\methods}[0]{Methods\xspace}

\DeclareSIUnit{\kcal}{kcal}
\DeclareSIUnit{\bohr}{Bohr}  

\newcommand*{\fref}[1]{Fig.~\ref{#1}}
\newcommand*{\tref}[1]{Table~\ref{#1}}
\newcommand*{\eref}[1]{Eq.~\eqref{#1}}
\newcommand*{\sref}[1]{Section~\ref{#1}}

\newcommand*{\olcite}[1]{Ref.~\cite{#1}}

\begin{document}

\title{Atomistic Machine Learning with Irreducible Cartesian Natural Tensors}

\author{Qun Chen}
\affiliation{Institute of Fundamental and Frontier Sciences, University of Electronic Science and Technology of China, Chengdu, China}

\author{A. S. L. Subrahmanyam Pattamatta}
\affiliation{Department of Mechanical Engineering, The University of Hong Kong, Hong Kong SAR, China}

\author{Boyu Wang}
\affiliation{Institute of Fundamental and Frontier Sciences, University of Electronic Science and Technology of China, Chengdu, China}

\author{David J. Srolovitz}
\affiliation{Department of Mechanical Engineering, The University of Hong Kong, Hong Kong SAR, China}

\author{Mingjian Wen}
\email{mjwen@uestc.edu.cn}
\affiliation{Institute of Fundamental and Frontier Sciences, University of Electronic Science and Technology of China, Chengdu, China}

\begin{abstract}

  Atomistic machine learning is a powerful tool for accurate and efficient investigation of material behavior at the atomic scale.
  While attempts have been made to construct models directly within Cartesian space, they face challenges in providing a systematic framework based on irreducible representations---a core feature of widely used spherical models.
  Here we propose Cartesian Natural Tensor Networks to overcome these limitations and thus offer a general, symmetry-preserving framework for atomistic machine learning.
  We present a theory of irreducible representations using Cartesian natural tensors, comprising their construction, their products, and a systematic scheme to decompose and reconstruct high-rank physical tensors.
  Leveraging this machinery, we develop equivariant machine learning interatomic potentials for materials and molecular systems with performance on par with leading spherical models.
  It further captures accurate structure--property relationships for tensorial quantities ranging from low-rank dipole moments to high-rank tensors with complex symmetries, such as the elastic constant tensor.

\end{abstract}

\maketitle

\section{Introduction}

Atomistic machine learning (ML) represents a data-driven paradigm that learns to predict material and molecular properties directly from the atomic structure. 
The primary inputs are the spatial coordinates of atoms and their chemical identities, while the outputs span interatomic potential energies and forces~\cite{unke2021machine,musil2021physics}; scalar properties such as bond dissociation energies and band gaps~\cite{wen2020bondnet,ruff2024connectivity}; and tensorial quantities including dipole moments, polarizabilities, and elastic constant tensors~\cite{veit2020predicting,wen2024an}.
When trained on quantum-mechanical data, these models approach ab initio accuracy at a fraction of the computational cost, enabling simulations of larger systems and longer time scales, and systematic exploration across vast chemical and materials spaces.
By uniting accurate physical data with scalable learning approaches, atomistic ML underpins transferable interatomic potentials, rapid property screening, and the construction of robust structure--property relationships.
Atomistic ML has become foundational in the sciences and engineering, accelerating rational discovery and design across diverse fields, including energy conversion and storage~\cite{yao2023machine,wen2023chemical}, heterogeneous catalysis~\cite{li2023data,mou2023bridging}, and pharmaceutical discovery~\cite{vamathevan2019applications, ekins2019exploiting}, $\ldots$

At the core of any atomistic ML model lies the central task of describing the local atomic environment around each atom.
This step is critical to any atomistic ML because it converts raw atomic coordinates and chemical identities into numerical representations that ML algorithms can process.
Early methods relied on descriptors designed to capture important structural features like bond lengths and bond angles (e.g., atom-centered symmetry functions~\cite{behler2007generalized}, the Coulomb matrix~\cite{rupp2012fast}, and DeePMD descriptors~\cite{zhang2018deep}).
More recently, the field has evolved toward more systematic approaches based on mathematical expansions of the atomic environment using well-established basis functions.
Current state-of-the-art approaches use two key mathematical tools: polynomials and/or trigonometric functions to describe distance relationships between atoms, and spherical harmonics to capture angular relationships.
Examples include SNAP~\cite{thompson2015spectral}, ACE~\cite{drautz2019atomic}, TFN~\cite{thomas2018tensor}, NequIP~\cite{batzner2022e}, MACE~\cite{batatia2022mace}, and GRACE~\cite{bochkarev2024graph}.
In other words, these methods first transform atomic structures from their Cartesian coordinates (i.e., \{x, y, z\} positions) into spherical coordinates (distances and angles), and then train the ML model in this spherical representation.
However, since atomic structures are naturally described in Cartesian coordinates for most simulation methods and property calculations, there is a compelling motivation to develop methodologies that work directly in Cartesian space, directly utilizing geometric information in its natural form while maintaining clear physical interpretability.

Considerable progress has been made in developing atomistic ML methods that operate directly in Cartesian space.
For example, models such as
MTP~\cite{shapeev2016moment},
REANN~\cite{zhang2021physically},
CACE~\cite{cheng2024cartesian},
CAMP~\cite{wen2025cartesian},
and
ICTP~\cite{zaverkin2024higher}
build on the ideas of Cartesian atomic moments or Cartesian cluster expansion to construct interatomic potentials.
FieldSchNet~\cite{gastegger2021machine},
TensorNet~\cite{simeon2023tensornet},
HotPP~\cite{wang2024equivariant},
and TACE~\cite{xu2025tace}
utilize Cartesian representations to predict low-rank tensorial properties.
Despite these achievements, Cartesian approaches still face key challenges that limit their broader applicability.
First, models employing reducible Cartesian tensors, such as HotPP~\cite{wang2024equivariant} and CAMP~\cite{wen2025cartesian}, naturally accommodate arbitrary-rank tensors and respect global rotational symmetry, but do not systematically enforce the internal index symmetries of physical tensors (e.g., $C_{ijkl} = C_{jikl} = C_{klij}$ of the elastic constant tensor), leaving unphysical redundancies in the predicted tensors.
Second, models employing irreducible Cartesian representations, such as TensorNet~\cite{simeon2023tensornet}, ICTP~\cite{zaverkin2024higher}, and TACE~\cite{xu2025tace}, could in principle be extended to higher ranks, but in practice remain limited to low-rank tensorial properties due to the absence of
a systematic scheme for the decomposition and reconstruction of physical tensors using irreducible Cartesian building blocks.
These limitations motivate a principled and systematic framework to fully harness the advantages of working directly in Cartesian space.

This work addresses these challenges by developing a comprehensive computational framework for atomistic ML using Cartesian natural tensors.
A natural tensor is one whose components are determined by the geometric properties of the space in which it lives and is, therefore, an intrinsic geometric construction rather than a regular tensor.
Natural tensors find wide application in fluid physics (e.g., Navier--Stokes equations)~\cite{kundu2024fluid}, electromagnetism (e.g., multipole expansions of charge distributions)~\cite{jones2013theory}, and increasingly in machine learning (e.g., equivariant neural networks)~\cite{zaverkin2024higher,simeon2023tensornet}.
In this work, we adapt the theory of irreducible Cartesian natural tensors for atomistic ML, further develop a systematic scheme for the decomposition and reconstruction of arbitrary physical tensors, and provide the mathematical tools and software needed to apply this framework to atomic systems.

Building on this foundation, we introduce a modeling framework applicable to both interatomic potentials and structure--property relationships of high-rank tensorial observables with intrinsic symmetries.
Our Cartesian Natural Tensor Networks (\model) is E(3)-equivariant; i.e., it respects the three-dimensional rotational, translational, and inversion symmetries inherent to atomic systems.
We demonstrate the capabilities of \model through a range of atomistic ML tasks.
As an MLIP, \model achieves accuracy comparable to leading spherical-tensor approaches across both materials and molecular systems, with universal models that generalize across the periodic table.
Beyond potentials, \model provides a unified route to predicting tensorial properties of varying rank and symmetry, including dipole moments, polarizabilities, nuclear magnetic shielding, and elastic constant tensors.
\model enables the exploration of the rich landscape of Cartesian representations and their applications in atomistic ML.

\section{Results}
\label{sec:results}

\subsection{Cartesian Natural Tensor Theory}
\label{sec:nat:tensor:theory}

\begin{figure*}[tbh!]
  \centering
  \includegraphics[width=0.7\linewidth]{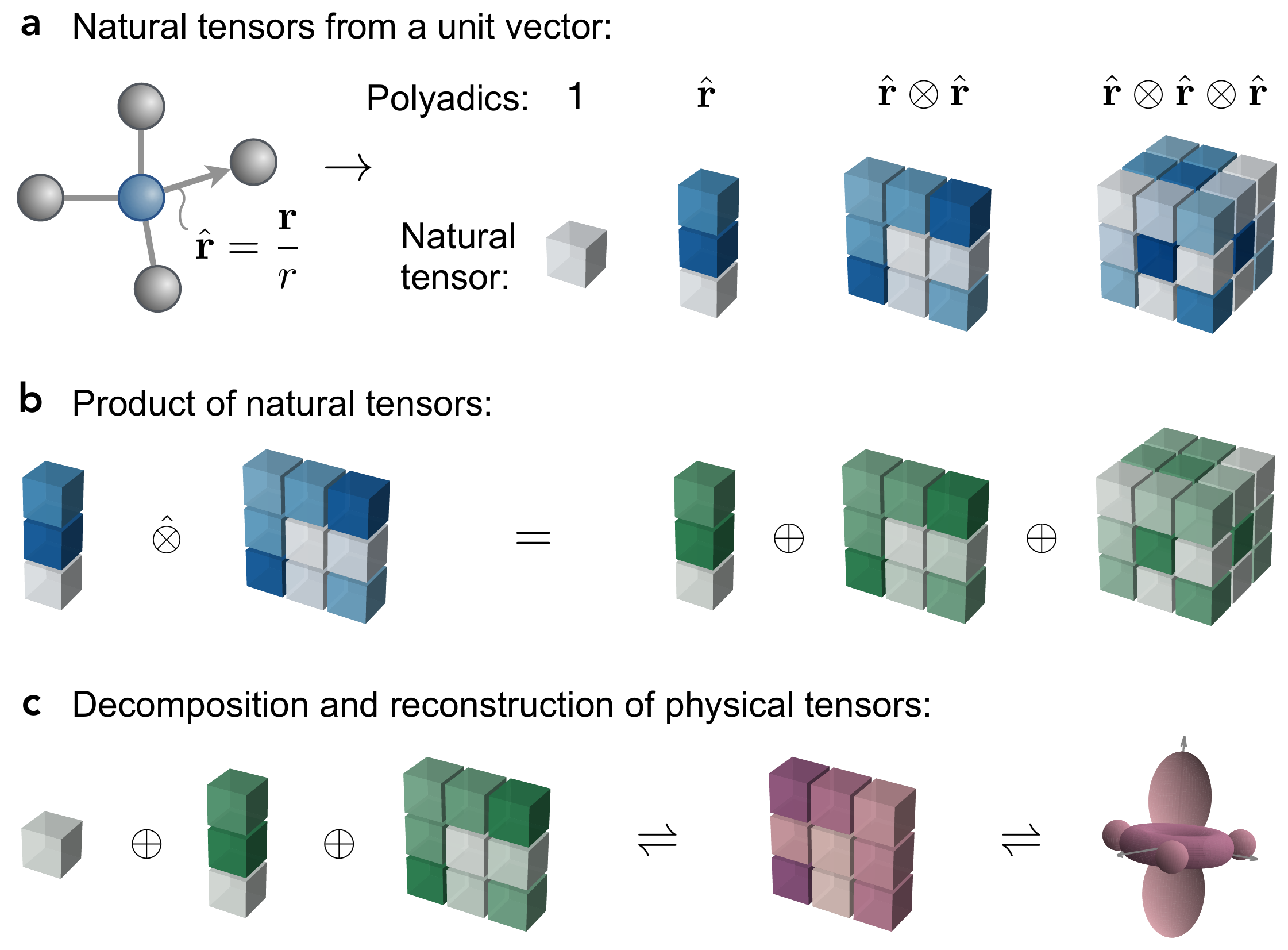}
  \caption{\textbf{Schematic illustration of Cartesian natural tensor operations.}
    \textbf{a}.\ Construction of natural tensors of different ranks from a unit vector $\hat{\mathbf{r}}$.
    \textbf{b}.\ Tensor product between a rank-1 and a rank-2 natural tensor generates three natural tensors of ranks 1, 2, and 3.
    \textbf{c}.\ Any physical tensor (e.g., the nuclear shielding tensor) can be decomposed into a set of natural tensors and, conversely, reconstructed from them.
    $\otimes$ is the product between ordinary tensors,
    $\hat\otimes$ is the  product between natural tensors, and
    $\oplus$ is the direct sum of natural tensors.
  }
  \label{fig:natural:tensor}
\end{figure*}

A natural tensor is a fully symmetric Cartesian tensor whose traces vanish on any pair of indices~\cite{coope1965irreducible,coope1970irreducible2,coope1970irreducible3}.
Formally, a rank-$n$ tensor $\mathbf{X}_n$ with components $X_{i_1i_2\dots i_n}$
is natural if it is
\begin{enumerate}[nosep]
  \item Symmetric: $X_{i_{\pi(1)}i_{\pi(2)}\dots i_{\pi(n)}} = X_{i_1i_2\dots i_n}$ for all $\pi \in S_n$, where $S_n$ is the symmetric group of degree $n$ (the group of all permutations of $n$ indices),
  \item Traceless: $\delta_{i_a i_b} X_{i_1\dots i_a\dots i_b \dots i_n} = 0$ for all $1\leq a < b \leq n$,
\end{enumerate}
where $\delta_{ij}$ is the Kronecker delta and Einstein summation is implied over repeated indices (the same below unless otherwise stated).
Scalars (rank-0 tensors) and vectors (rank-1 tensors) are trivially natural tensors.

Natural tensors are widely used to represent physical properties.
For example, any rank-2 Cartesian tensor $\mathbf{T}$ ($T_{ij}$) can be decomposed into
\begin{equation} \label{eq:decomp:rank2}
  T_{ij}=
  \underbrace{\frac{1}{3}T_{kk}\delta_{ij}}_{n=0}
  + \underbrace{\frac{1}{2}\left(T_{ij}-T_{ji}\right)}_{n=1}
  + \underbrace{\frac{1}{2}\left(T_{ij}+T_{ji}\right)-\frac{1}{3}T_{kk}\delta_{ij}}_{n=2},
\end{equation}
where the $n=0,1,2$ natural tensors are typically called the isotropic, antisymmetric, and symmetric parts, respectively (\fref{fig:natural:tensor}c).
This decomposition is useful because each part has a distinct physical meaning.
For example, if $\mathbf{T}$ is a stress tensor, the isotropic part represents the hydrostatic pressure (causing volumetric change), the antisymmetric part is related to torques (zero in classical continuum mechanics), and the symmetric part represents the shear stress (causing shape change).
Conversely, given three natural tensors of ranks 0, 1, and 2, one can fully reconstruct a rank-2 Cartesian tensor.

In three dimensions, a rank-$n$ natural tensor $\mathbf{X}_n$ has $3^n$ components, but only $2n+1$ of these are independent (due to the symmetry and traceless constraints).
A rank-$n$ natural tensor $\mathbf{X}_n$ furnishes a $2n+1$-dimensional irreducible representation of the special orthogonal group SO(3) (the group of 3D rotations).
Intuitively, under rotations, the $2n+1$ independent components mix in a way that prevents decomposition into smaller sets of components that transform independently under SO(3);
formally, the representation space of $\mathbf{X}_n$ is irreducible under SO(3), meaning it contains no proper nontrivial SO(3)-invariant subspaces~\cite{zee2016group}.
This  makes natural tensors particularly well-suited for atomistic ML, where rotational equivariance is crucial.
Below, we summarize the three fundamental operations on natural tensors that serve as the basis for our framework (a detailed description is provided in Supplementary Note~1).

\textbf{Natural tensors from a unit vector}.
Given a unit vector $\hat{\mathbf{r}}$, a rank-$n$ natural tensor $\mathbf{X}_n$ can be constructed in two steps (\fref{fig:natural:tensor}a).
1.\ Symmetric polyadic tensor: Form the rank-$n$ tensor $\mathbf{U} = \mathbf{r}^{\otimes n} = \hat{\mathbf{r}} \otimes \hat{\mathbf{r}} \otimes \dots \otimes \hat{\mathbf{r}}$ (repeated $n$ times).
By construction, $\mathbf{U}$ is fully symmetric.
2.\ Trace removal: Project $\mathbf{U}$ into the symmetric traceless subspace via $\mathbf{X}_n  = \mathbf{H} \odot^n  \mathbf{U}$,
where $\mathbf{H}$ is the rank-$2n$ projection tensor, and
$\odot^n$ denotes $n$-fold contraction over $n$ pairs of indices (i.e., $X_{k_1 \dots k_n} = H_{k_1 \dots k_n i_1 \dots i_n} U_{i_1 \dots i_n}$).
$\mathbf{H}$ is given by~\cite{jerphagnon1978description}
\begin{equation} \label{eq:H:unit:vector}
  H_{k_1\dots k_n i_1\dots i_n} = \sum_{t=0}^{\lfloor n/2 \rfloor} (-1)^t \cdot F \cdot \{\delta_{ki}^{\otimes{n-2t}}\delta_{kk}^{\otimes t} \} \delta_{ii}^{\otimes t},
\end{equation}
where
$ F = \frac{(2n-2t-1)!!}{(2n-1)!!}$,
the superscript $^{\otimes t}$ is a shorthand notation for the $t$-fold tensor product of Kronecker deltas (e.g., $ \delta_{ki}^{\otimes 2} = \delta_{k_1i_1}\delta_{k_2i_2}$ and $\delta_{ii}^{\otimes 2} = \delta_{i_1i_2}\delta_{i_3i_4}$),
and the curly braces $\{\,\}$ denote symmetrization achieved by summing over all unique permutations of the indices.
For example, when $n=2$, we have
$
  H_{k_1k_2i_1i_2} = \delta_{k_1 i_1} \delta_{k_2 i_2}  - \frac{1}{3} \delta_{k_1 k_2} \delta_{i_1 i_2},
$
and the corresponding rank-2 natural tensor is
$
  X_{k_1k_2}
  = H_{k_1k_2i_1i_2} U_{i_1i_2}
  = r_{k_1}r_{k_2} - \frac{1}{3} \delta_{k_1 k_2} r_{i_1} r_{i_1}.
$
The projection tensor $\mathbf{H}$ is analogous to spherical harmonics that generate irreducible representations of SO(3) from a unit vector, but it is expressed entirely in Cartesian form.

\textbf{Product of natural tensors}.
The tensor product between two natural tensors can be expressed as a direct sum of a set of natural tensors~\cite{coope1970irreducible3,lehman1989angular}.
Given natural tensors $\mathbf{X}_{l_1}$ of rank $l_1$ and $\mathbf{Y}_{l_2}$ of rank $l_2$, their product
\begin{equation} \label{eq:tp}
  \mathbf{Z}_{l_3} = \mathbf{X}_{l_1} \hat\otimes \mathbf{Y}_{l_2}
\end{equation}
is a tensor whose rank lies in the range $|l_1 - l_2| \leq l_3 \leq l_1 + l_2$ (analogous to spherical tensors~\cite{edmonds1996angular}), where $\hat\otimes$ represents the natural tensor product (e.g., when $l_1=1$ and $l_2=2$, $\mathbf{Z}_{l_3}$ is of rank 1, 2, or 3; see \fref{fig:natural:tensor}b).
To obtain $\mathbf{Z}_{l_3}$, we can first compute the ordinary tensor product $\mathbf{W} = \mathbf{X}_{l_1} \otimes \mathbf{Y}_{l_2}$, and then symmetrize $\mathbf{W}$ and remove the traces.
Similar to the construction of natural tensors from unit vectors, based on the formulas in Ref.~\cite{lehman1989angular}, we have derived an explicit expression for a rank-$(l_1+l_2+l_3)$ projection tensor $\mathbf{H}$ that performs this operation.
For even $l_1 + l_2 - l_3=2d$, we have
\begin{equation} \label{eq:tp:even:H}
  \begin{aligned}
     & H_{k_1\dots k_{l_3} i_1\dots i_{l_1} j_1\dots j_{l_2}} = \\
     & \sum_{t=0}^{\min(l_1,l_2)-d} (-2)^t
    \cdot F \cdot
    \{ \delta_{ik}^{\otimes l_1 - (d+t)} \delta_{jk}^{\otimes l_2 - (d+t)} \delta_{kk}^{\otimes t} \} \delta_{ij}^{\otimes d+t} ,
  \end{aligned}
\end{equation}
where $F = \frac{(2l_3-2t-1)!!}{(2l_3-1)!!}$.
For odd $l_1 + l_2 - l_3=2d+1$, the formula is similar but with the multiplication of an additional Levi--Civita symbol (expression given in Supplementary Note~1).
With this, $\mathbf{Z}_{l_3}$ can be computed as
$\mathbf{Z}_{l_3} = \mathbf{H} \odot^{l_1} \mathbf{X}_{l_1} \odot^{l_2} \mathbf{Y}_{l_2}$
(i.e.,
$
  Z_{k_1\dots k_{l_3}} = H_{k_1\dots k_{l_3} i_1\dots i_{l_1} j_1\dots j_{l_2}} X_{i_1\dots i_{l_1}} Y_{j_1\dots j_{l_2}}
$
).
Here, the projection tensor $\mathbf{H}$ is analogous to the Clebsch--Gordan coefficients~\cite{edmonds1996angular} for spherical tensor product.

\textbf{Decomposition and reconstruction of physical tensors}.
A central task in utilizing natural tensors involves converting between regular Cartesian tensors and their natural representations.
The decomposition and reconstruction in \eref{eq:decomp:rank2} for rank-2 tensors are possible for arbitrary tensors; however, the complexity increases significantly with the rank and internal index symmetries of the tensor.
Unlike the rank-2 case in \eref{eq:decomp:rank2}, higher-rank tensors yield multiple linearly dependent decomposition candidates~\cite{coope1970irreducible2}, which should be carefully selected to form a linearly independent basis set of natural tensors.
Moreover, the internal index symmetries of a physical tensor impose additional constraints on the form of natural tensors~\cite{jerphagnon1978description}.
For example, a generic rank-4 tensor yields six rank-3 natural tensor candidates in its decomposition spectrum, but only three of them are linearly independent.
The rank-4 elastic constant tensor $\mathbf{C}$ has symmetries $C_{ijkl} = C_{jikl} = C_{klij}$, under which, all associated rank-3 candidates vanish.
Several studies have examined the decomposition of physical tensors with specific ranks~\cite{morris1969averaging,harris1974theory, jerphagnon1978description,russ1982properties}; a more general approach was proposed by Coope et al~\cite{coope1965irreducible,coope1970irreducible2,coope1970irreducible3}.
However, a general method for tensors of arbitrary rank and symmetry remains unavailable (to our knowledge).
We have developed a systematic approach to addressing the challenges, which consists of three key steps.

First, create the rank-$(m+n)$ tensor $\mathbf{G}$ to map between the rank-$n$ physical tensor space and the rank-$m$ natural tensor space~\cite{coope1970irreducible2}:
\begin{equation} \label{eq:natural:projector}
  G_{k_1\dots k_m i_1\dots i_n}  = \sum_{t=0}^{\lfloor m/2 \rfloor} c_t \{\{\delta_{ki}^{\otimes(m-2t)} \delta_{kk}^{\otimes t} \delta_{ii}^{\otimes t}
  \}\}
  \delta_{ii}^{\otimes (n-m)/2},
\end{equation}
where $c_t$ are known coefficients (given in Supplementary Note~1),
the double curly braces $\{\{\,\}\}$ denote symmetrization achieved by averaging over all unique permutations of the indices.
Depending on the assignment of the indices, there exist multiple candidate mapping tensors for a given $n$ and $m$.
For example, when $n=3$ and $m=1$, we have
$
  G_{k_1 i_1 i_2 i_3}
  =  c_0 \delta_{ki}  \delta_{ii}
$
(no Einstein summation on $i$),
resulting in three candidates:
$
  G^1  =  c_0 \delta_{k_1i_1} \delta_{i_2i_3},
$
$
  G^2  =  c_0 \delta_{k_1i_2} \delta_{i_3i_1},
$
and
$
  G^3  =  c_0 \delta_{k_1i_3} \delta_{i_1i_2},
$
where the subscript $k_1i_1i_2i_3$ is dropped for clarity.

Next, identify a subset of $N_g$ unique linearly independent mapping tensors from all $N$ candidates $\mathbf{G}^1, \ldots, \mathbf{G}^N$.
This can be achieved via a QR factorization~\cite{golub1996matrix}, followed by symmetry-informed elimination according to the intrinsic symmetries of the physical tensor.
For example, when $n=3$ and $m=1$, a QR factorization will verify that all three candidates above are linearly independent.
But for a rank-3 tensor with symmetry $T_{i_1i_2i_3} = T_{i_1i_3i_2}$, $\mathbf{G}^2$ and $\mathbf{G}^3$ yield identical outcomes upon contraction with $\mathbf{T}$; thus, only $\mathbf{G}^1$ and $\mathbf{G}^2$ are retained as unique mapping tensors.

Finally, construct the decomposition projector $\mathbf{H}$ and embedding tensor $\mathbf{Q}$ to map between a physical tensor with arbitrary rank and symmetry and its natural representation:
\begin{equation} \label{eq:H:G}
  \begin{aligned}
    \mathbf{H}^p_{m+n} & = \sum_{q=1}^N h_{pq} \mathbf{G}^q_{m+n}                                 \\
    \mathbf{Q}^p_{m+n} & = \mathbf{G}^p_{m+n} + \sum_{q=N_g+1}^{N} \beta_{qp} \mathbf{G}^q_{m+n},
  \end{aligned}
\end{equation}
where $p=1,2,\dots,N_g$, and $h_{pq}$ and $\beta_{qp}$ (expressions given in Supplementary Note~1) are coefficients that can be determined by enforcing the orthonormal condition between different $\mathbf{G}^q$.
With them, one can utilize
$
  \mathbf{X}^p_m = \mathbf{H}^p_{m+n} \odot^n \mathbf{T}_n
$
to extract the natural tensors from an ordinary tensor,
and, conversely, utilize
\begin{equation} \label{eq:reconstruct}
  \mathbf{T}_n = \sum_{m=0}^n\sum_{p=1}^{N_g} \mathbf{Q}^p_{m+n} \odot^m \mathbf{X}^p_m
\end{equation}
to reconstruct an ordinary tensor from its natural representations.

\textbf{Summary}.
The contributions of our theoretical developments are twofold.
First, we have developed a systematic approach to decomposing and reconstructing physical tensors of arbitrary rank and symmetry using natural tensors, which is not available in the literature (to our knowledge).
Second, we have derived explicit expressions for the projection tensors (Eqs.~\eqref{eq:H:unit:vector}, \eqref{eq:tp:even:H}, and \eqref{eq:H:G}) that consist of only Kronecker deltas and Levi--Civita symbols, which can be precomputed and stored, enabling efficient natural tensor operations.

\subsection{The \model Model}

\begin{figure*}[tbh!]
  \centering
  \includegraphics[width=0.8\linewidth]{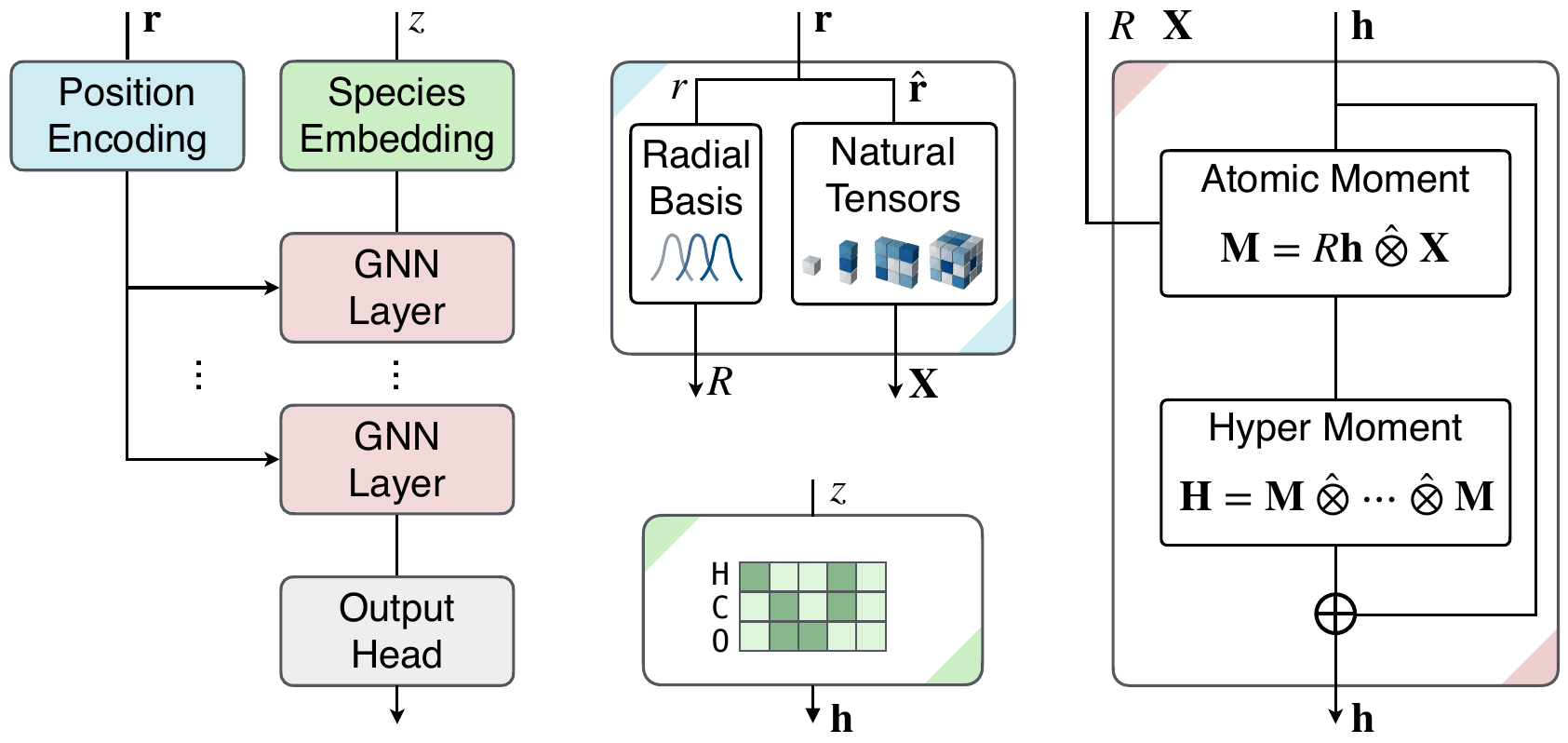}
  \caption{\textbf{Overview of the \model model architecture.}
    The relative distance vector $\mathbf{r}$ of an atom from its neighbor is encoded using a set of radial basis functions for its magnitude $r$ and natural tensors constructed from polyadics of the unit vector $\hat{\mathbf{r}}$.
    The atomic species $z$ is encoded using a learnable embedding to generate the initial atom features $\mathbf{h}$.
    With the radial part $R$, the angular part $\mathbf{X}$,
    and the atom features $\mathbf{h}$, each graph neural network (GNN) layer first constructs the atomic moment $\mathbf M$ and then the hyper moment $\mathbf H$ using natural tensor products (denoted by $\hat\otimes$).
    Finally, the atomic features are mapped to the target properties using an output head.
  }
  \label{fig:model}
\end{figure*}

Leveraging the three fundamental operations of natural tensors and the concept of moment tensors as introduced in MTP~\cite{shapeev2016moment} and CAMP~\cite{wen2025cartesian}, we propose \model: a theoretically grounded framework for constructing equivariant atomistic ML models.
\model takes an atomic structure as input, generates equivariant atom features, and produces interatomic potentials or structure--property relationships, including those involving high-rank tensors.
A schematic overview of the model architecture is illustrated in \fref{fig:model}.

The model begins by encoding the atomic structure.
The relative position vector $\mathbf{r}$ between an atom and its neighbor is decomposed into its scalar distance $r$ and the unit directional vector $\hat{\mathbf{r}} = \mathbf{r}/r$.
The scalar distance $r$ is then expanded into $R$ using a set of radial basis, specifically Chebyshev polynomials of the first kind.
Angular information is incorporated through $\hat{\mathbf{r}}$, which is transformed into natural tensors $\mathbf{X}$ via the first operation described in \sref{sec:nat:tensor:theory}.
Atomic numbers $z$ are embedded into learnable initial atom features $\mathbf{h}$, completing the initial encoding of the atomic structure.

The core architecture of \model comprises a multilayer graph neural network (GNN) that iteratively refines atom features.
Within each GNN layer, the atomic moment $\mathbf{M}$ is 
constructed as a tensor product involving radial function $R$, current atom features $\mathbf{h}$, and the angular component $\mathbf{X}$.
Subsequently, a self-tensor product is performed on $\mathbf{M}$, yielding the hyper moment $\mathbf{H}$ that encodes many-body interactions~\cite{shapeev2016moment,drautz2019atomic,batatia2022mace}.
Both steps employ the natural tensor product formalism introduced above, 
ensuring that $\mathbf{M}$ and $\mathbf{H}$ are natural tensors.
Empirical results (below) indicate that typically two to three GNN layers suffice to attain high predictive accuracy.

At the final stage, relevant natural tensors derived from atom features $\mathbf{h}$ are selected to construct the desired physical quantities.
For interatomic potentials, this process involves extracting the rank-0 natural tensor (scalar).
For higher-rank tensorial properties, the relevant natural tensors are extracted from atom features according to their decomposition and reconstruction spectrum, and the target physical tensor is reconstructed from these components using the third operation in \sref{sec:nat:tensor:theory}.

The overall model architecture, the reconstruction of atomic and structural physical tensors, and the training procedures are provided in the \methods section.
A detailed comparison between \model and existing Cartesian methods for atomistic ML (e.g.,
TensorNet~\cite{simeon2023tensornet},
HotPP~\cite{wang2024equivariant},
ICTP~\cite{zaverkin2024higher},
and TACE~\cite{xu2025tace}) is given in Supplementary Note~2.

\subsection{Bulk LiPS and Water}
\label{sec:bulk system}

\begin{table}[tbh!]
  \centering
  \caption{\textbf{Model performance on the LiPS and Water datasets}.
    For LiPS, mean absolute errors (MAEs) of energy and forces on the test set are reported.
    For water, root-mean-square errors (RMSEs) of energy and forces are reported.
    CAMP results are from \olcite{wen2025cartesian}, CACE and others are from \olcite{cheng2024cartesian}.
  }
  \label{tab:lips:water}
  \begin{tabular}{llcc}
    \hline
    Dataset                 & Model                             & Energy (meV)  & Forces (meV/\AA) \\
    \hline
    \multirow{4}{*}{LiPS}   & NequIP~\cite{batzner2022e}        & 0.12          & 7.7              \\
                            & CAMP~\cite{wen2025cartesian}      & 0.12          & 7.4              \\
                            & \model (2 layers)                 & 0.11          & 6.4              \\
                            & \model (3 layers)                 & \textbf{0.09} & \textbf{5.6}     \\
    \hline
    \multirow{10}{*}{Water} & BPNN~\cite{behler2007generalized} & 2.3           & 120              \\
                            & ACE~\cite{drautz2019atomic}       & 1.7           & 99               \\
                            & REANN~\cite{zhang2021physically}  & 0.8           & 53               \\
                            & DeePMD~\cite{zhang2018deep}       & 2.1           & 92               \\
                            & NequIP~\cite{batzner2022e}        & 0.94          & 45               \\
                            & MACE~\cite{batatia2022mace}       & 0.63          & 36               \\
                            & CACE~\cite{cheng2024cartesian}    & 0.59          & 47               \\
                            & CAMP~\cite{wen2025cartesian}      & 0.59          & 34               \\
                            & \model (2 layers)                 & \textbf{0.54} & 34               \\
                            & \model (3 layers)                 & \textbf{0.54} & \textbf{31}      \\
    \hline
  \end{tabular}
\end{table}

\begin{figure*}[tbh!]
  \centering
  \includegraphics[width=0.95\linewidth]{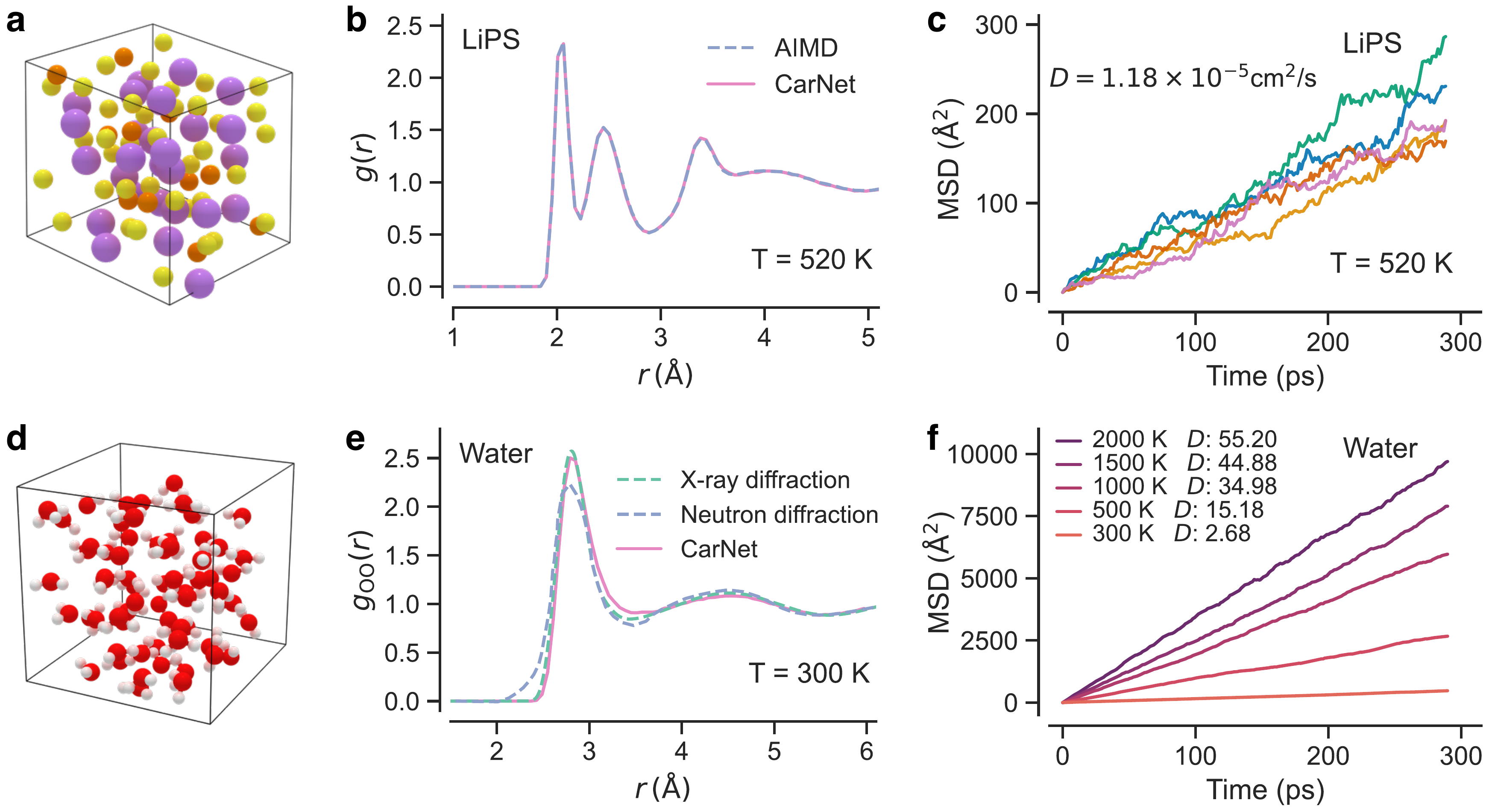}
  \caption{\textbf{Molecular dynamics (MD) simulation results of bulk LiPS and water systems.}
    \textbf{a}--\textbf{c}: Crystal structure, radial distribution function (RDF), and mean squared displacement (MSD) of Li$^+$ ions versus time of LiPS.
    \textbf{d}--\textbf{f}: Simulation cell, RDF of oxygen-oxygen pairs, and MSD versus time of bulk water.
    Five separate MD simulations with different initial velocities were performed for LiPS, and the reported diffusion coefficient $D$ in panel c is the average over the five runs.
    The reference ab initio molecular dynamics (AIMD) and experimental results are at the same temperatures shown in the figures, except for the X-ray diffraction data, which is at 295~K.
    In panel c, five MD simulations with different initial velocities were performed, and the average diffusion coefficient $D$ is reported.
    The MD simulation for water used a $2\times2\times2$ replication of this unit cell shown in panel d.
    In panel f, $D$ is given in the units of $\times10^{-5}$\si{\cm^2\per\s}.
    Simulation cells are plotted using AtomViz~\cite{atomviz};
    atom colors: purple (Li), orange (P), yellow (S), red (O), and white (H).
  }
  \label{fig:bulk:result}
\end{figure*}

We first apply \model to develop interatomic potentials for periodic systems---specifically, inorganic lithium phosphorus sulfide (LiPS) (a solid-state electrolyte)~\cite{batzner2022e} and bulk liquid water~\cite{cheng2019ab}.
We assess the model's performance by computing the mean absolute error (MAE) or root mean square error (RMSE) of energies and atomic forces on the test sets, benchmarking against state-of-the-art models in the literature, including both spherical models (e.g., NequIP~\cite{batzner2022e} and MACE~\cite{batatia2022mace}) and Cartesian models (e.g., CACE~\cite{cheng2024cartesian} and CAMP~\cite{wen2025cartesian}).
For both systems, \model configured with two GNN layers achieves the lowest reported errors, as shown in Table~\ref{tab:lips:water}; increasing to three layers further reduces errors.

Beyond low MAEs, \model enables stable molecular simulations to accurately predict structural and dynamical properties.
While low errors in energy and forces are necessary, they are not sufficient to guarantee the physical reliability of molecular dynamics (MD) simulations; unphysical force predictions can lead to instability or drift during simulations despite good MAEs~\cite{fu2023forces}.

To evaluate the stability and physical fidelity, we first perform MD simulations for LiPS under an NVT ensemble.
For LiPS, five MD simulations with different initial velocities were carried out using a timestep of 1~fs over a total duration of 300~ps (simulation details in \methods).
No instabilities or trajectory collapse were observed for all MD runs.
We next analyze the radial distribution functions (RDF) derived from the MD trajectories.
The RDF (\fref{fig:bulk:result}b) predicted by \model closely matches AIMD simulation results~\cite{cheng2019ab}.
We also compute the diffusion coefficient from the mean squared displacement (MSD) of the MD trajectories.
The MSD of \ce{Li+} in LiPS (\fref{fig:bulk:result}c) exhibits linear behavior, indicative of normal diffusion.
The calculated diffusion coefficient from the five MD runs is $D = (1.18\pm0.19) \times10^{-5}$~\si{\cm^2\per\s}, consistent with AIMD estimates of $D = 1.37 \times10^{-5}$~\si{\cm^2\per\s}~\cite{batzner2022e}.

We also conducted MD simulations for water using similar setups as those for LiPS (simulation details in \methods).
The RDF of oxygen-oxygen pairs in water at 300~K (\fref{fig:bulk:result}e) agrees well with X-ray~\cite{skinner2014the} and neutron~\cite{soper1997site} diffraction data.
The predicted diffusion coefficient is $D = 2.68 \times10^{-5}$~\si{\cm^2\per\s} at 300~K
as compared with AIMD results of $D = 2.67 \times10^{-5}$~\si{\cm^2\per\s}~\cite{cheng2019ab}.
To further check the stability of \model,
following~\cite{cheng2024cartesian}, we performed MD simulations at high temperatures up to 2000~K.
At high temperatures, the MD simulations remained stable, and the MSD exhibited linear behavior (\fref{fig:bulk:result}f).
This demonstrates the superior stability of the \model.
We do not expect physical quantities (e.g., $D$ in \fref{fig:bulk:result}f and RDF in Supplementary Figure~1) at high temperatures to be quantitatively accurate; this example serves only to demonstrate the stability of \model.

\subsection{Molecular Ethanol}

\begin{table*}[htb!]
  \centering
  \caption{\textbf{Prediction errors in ethanol properties}.
    Mean absolute errors (MAEs) on the test set are reported for energy $\mathcal{E}$, forces $\mathbf{F}$, dipole moment $\bm\mu$, polarizability $\bm\alpha$, nuclear chemical shift $\delta_{\text{all}}$ and shielding tensor $\bm\sigma_\text{all}$ for all atoms.
    $L$ is the maximum rank of the natural tensors employed.
  }
  \label{tab:ethanol:vacuum}
  \begin{tabular}{lcccccc}
    \hline
                                            & $\mathcal{E}$ (\si{\kcal\per\mol})
                                            & $\mathbf{F}$ (\si{\kcal\per\mol\per\angstrom})
                                            & $\bm\mu$ (D)
                                            & $\bm\alpha$ (\si{\bohr\cubed})
                                            & $\delta_{\text{all}}$ (ppm)
                                            & $\bm\sigma_\text{all}$ (ppm)                                                                                                          \\
    \hline
    PaiNN~\cite{schutt2021equivariant}      & 0.027                                          & 0.150          & 0.003           & 0.009           & -              & -              \\
    FieldSchNet~\cite{gastegger2021machine} & 0.017                                          & 0.128          & 0.004           & 0.008           & 0.169          & -              \\
    TensorNet~\cite{simeon2023tensornet}    & 0.008                                          & 0.058          & 0.003           & 0.007           & 0.139          & -              \\
    \model (multitask)                      & 0.0063                                         & 0.034          & 0.0011          & \textbf{0.0051} & 0.044          & 0.065          \\
    \model ($L=2$)                          & 0.0054                                         & 0.027          & 0.0009          & 0.0076          & 0.037          & 0.057          \\
    \model ($L=3$)                          & \textbf{0.0048}                                & \textbf{0.022} & \textbf{0.0007} & 0.0063          & \textbf{0.031} & \textbf{0.047} \\
    \hline
  \end{tabular}
\end{table*}

We also evaluated the performance of \model for ethanol as a small-molecule test.
In addition to energies and atomic forces, the ethanol dataset~\cite{gastegger2021machine} includes three tensorial properties: dipole moment $\bm\mu$, polarizability $\bm\alpha$, and nuclear shielding tensor $\bm\sigma$.
The dipole moment $\bm{\mu}$ is a rank-1 structural tensor defined at the molecular level, representing a property of the entire molecule.
The polarizability $\bm{\alpha}$ is a rank-2 structural tensor, also defined at the molecular level.
In contrast, the nuclear shielding $\bm{\sigma}$ is a rank-2 atomic tensor; i.e., each atom in the molecule has a  shielding tensor that is sensitive to the local electronic environment.
The nuclear chemical shift is a scalar property derived from $\bm\sigma$ as $\delta = \text{Tr}(\bm\sigma)/3$.

\model demonstrates significant improvements predicting these properties relative to existing state-of-the-art models such as FieldSchNet~\cite{gastegger2021machine} and TensorNet~\cite{simeon2023tensornet}.
Using a multitask learning framework with a shared backbone and multiple output heads to concurrently learn all properties, \model achieves the highest accuracy across all tasks (\tref{tab:ethanol:vacuum}).
For instance, the errors in predicting the dipole moment $\bm\mu$ and nuclear chemical shift $\delta_\text{all}$ are approximately $3\times$ smaller than those from TensorNet. 
Separate errors $\delta_\text{H}$, $\delta_\text{C}$, and $\delta_\text{O}$ for each element are given in Supplementary Note~3.
This highlights the model's ability to learn a unified representation in predicting distinct scalars, atomic tensors, and structural tensors.
We also trained models where each property was learned individually (energy and forces jointly), observing reductions in errors across all outputs except $\bm\alpha$.
Models employing natural tensor representations with a maximum rank of $L=3$ outperform their $L=2$ counterparts, underscoring the critical role of high-rank tensors in capturing complex interatomic interactions, while highlighting limitations of models that rely on tensor representations up to rank-1 and rank-2 (e.g., FieldSchNet and TensorNet).

\subsection{Crystal Elastic Constant Tensor}
\label{sec:elastic}

\begin{figure*}[tbh!]
  \centering
  \includegraphics[width=1.0\linewidth]{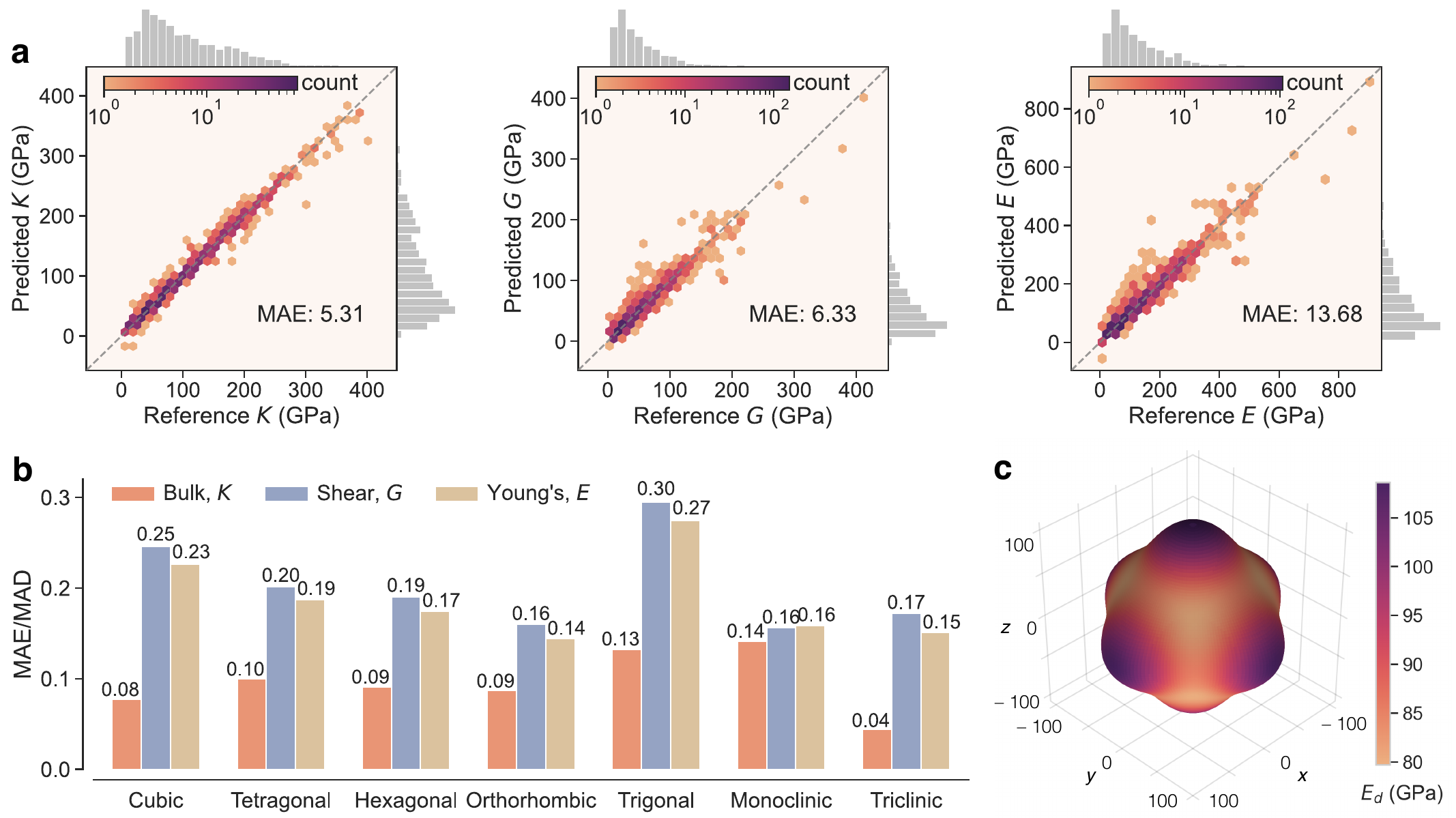}
  \caption{\textbf{Performance of \model in predicting elastic properties}.
    \textbf{a}.\ Predicted bulk modulus $K$, shear modulus $G$, and Young's modulus $E$ compared with reference density functional theory (DFT) values (84 elements).
    \textbf{b}.\ Normalized error by crystal system.
    \textbf{c}.\ Directional Young's modulus $E_d$ of CaS predicted by the model.
    The cubic symmetry of rocksalt CaS is clearly reflected in the predicted $E_d$.
    MAE is the mean absolute error, and MAD is the mean absolute deviation.
  }
  \label{fig:elastic:tensors}
\end{figure*}

We now demonstrate the capability of \model to predict the elastic constant tensor (characterizing linear elastic response of a material under applied stress) of inorganic crystalline solids.
Currently, no Cartesian atomistic ML methods can model the full (rank-4) elastic constant tensor (with up to 21 independent components depending on crystal symmetry).
This is more challenging due to both the inherent complexity of the elastic constant tensor and the compositional diversity of the dataset.
The dataset~\cite{wen2024an} encompasses structures spanning all seven crystal systems, involving 84 chemical elements, thereby presenting substantial variability in symmetry, composition, and structural complexity.
\model effectively handles this complexity, exhibiting robust applicability across diverse crystal symmetries and compositional variations.

\model directly predicts the full elastic constant tensor while inherently satisfying the two fundamental physical constraints: (1) frame indifference (coordinate system invariance) and (2) crystal symmetry adherence.
Frame indifference ensures the predicted tensor is equivariant under rigid rotations of the coordinate system, while the symmetry constraint guarantees that the tensor reflects the intrinsic point-group symmetries of the crystal~\cite{wen2024an}.
From the predicted rank-4 elastic constant tensor, we can obtain the Voigt stiffness matrix $C$~\cite{voigt1910lehrbuch}.
Scalar elastic moduli such as the bulk modulus $K$, shear modulus $G$, and Young's modulus $E$ are derived from $C$ using the Hill averaging scheme~\cite{hill1952elastic}.
Quantitatively, \model achieves MAEs of 5.31 for $K$, 6.39 for $G$, and 13.68 for $E$, over the 84 elements in the dataset (where values span from near 0--400~GPa for $K$ and $G$, and up to $\sim$800~GPa for $E$, - see \fref{fig:elastic:tensors}a).
The errors are $\sim$18\% lower than those by MatTen~\cite{wen2024an} and are comparable to those by XPaiNN~\cite{yan2025general}, both of which are spherical models specifically designed for elastic constant tensors (\tref{tab:mae:mad}).

\begin{table}[htb!]
  \caption{\textbf{Prediction errors in elastic properties.}
    Mean absolute errors (MAEs) of the bulk ($K$), shear ($G$), and Young's ($E$) moduli, as well as the $6\times6$ Voigt matrix ($C$).
    MAE of $C$ is calculated component-wise.
  }
  \label{tab:mae:mad}
  \centering
  \begin{tabular}{lcccc}
    \hline
                                             & $K$ (GPa)     & $G$ (GPa)     & $E$ (GPa)      & $C$ (GPa)     \\
    \hline
    AutoMatminer~\cite{dunn2020benchmarking} & 9.84          & 9.27          & 22.10          & -             \\
    MatSca~\cite{wen2024an}                  & 7.32          & 8.63          & 19.87          & -             \\
    MatTen~\cite{wen2024an}                  & 7.37          & 8.38          & 20.59          & 4.52          \\
    XPaiNN~\cite{yan2025general}             & 5.40          & \textbf{6.33} & 14.74          & -             \\
    \model                                   & \textbf{5.31} & 6.39          & \textbf{13.68} & \textbf{3.32} \\
    \hline
  \end{tabular}
\end{table}

A detailed analysis reveals that \model maintains consistent predictive accuracy across different crystal systems.
The relative error, MAE normalized by the mean absolute deviation (MAD) of the reference values, is comparable among all seven crystal systems (\fref{fig:elastic:tensors}b).
Despite dataset imbalance (see Supplementary Figure~2) favoring high-symmetry over low-symmetry crystals (e.g., cubic vs.\ triclinic),
\model demonstrates effective transferability and generalization, indicating its ability to learn unified representations 
across a broad structural spectrum.

Beyond scalar moduli $K$, $G$, and $E$, the ability to predict the full elastic constant tensor enables efficient analysis of anisotropic elastic behaviors.
For example, the directional Young's modulus $E_d$ can be computed for all orientations, enabling comprehensive evaluation of elastic anisotropy and directional stiffness variations (\fref{fig:elastic:tensors}c).
This capability underscores the utility of \model for materials design and mechanical property optimization where directional elastic responses are critical.

\subsection{Universal MLIP for Materials}

\begin{table}[tbh!]
  \caption{\textbf{MAEs for universal MLIPs on the MatPES r$^2$SCAN test set}.
    Model size is measured by the number of parameters in millions.
    Bold and underlined text indicate the smallest and second smallest errors, respectively.
    MACE and UPET are developed using additional data in addition to MatPES; both are pretrained on the OMat24 dataset and then finetuned on MatPES.
    M3GNet, CHGNet, and TensorNet results are from \olcite{kaplan2025matpes}; MACE and UPET results are evaluated by us using the model checkpoints released by the authors (see Data Availability).
  }
  \label{tab:matpes}
  \begin{tabular}{lcccc}
    \hline
                                         & Energy         & Forces         & Stress            & Model size \\
                                         & (meV/atom)     & (meV/\AA)      & (GPa)             & (millions) \\
    \hline
    M3GNet~\cite{chen2022a}              & 44             & 210            & 0.970             & 0.66       \\
    CHGNet~\cite{deng2023chgnet}         & 30             & 156            & 0.735             & 2.70       \\
    TensorNet~\cite{simeon2023tensornet} & 34             & 163            & 0.754             & 0.84       \\
    MACE~\cite{batatia2025fondation}     & 25             & \underline{84} & 0.711             & 9.06       \\
    UPET~\cite{mazitov2025petmad}        & \textbf{12}    & \textbf{40}    & \textbf{0.201}    & 192.89     \\
    \model (1 layer)                     & 39             & 175            & 0.897             & 1.60       \\
    \model (2 layers)                    & 27             & 141            & 0.717             & 3.30       \\
    \model (3 layers)                    & \underline{23} & 130            & \underline{0.642} & 7.80       \\
    \hline
  \end{tabular}
\end{table}

\begin{figure*}
  \includegraphics[width=1\linewidth]{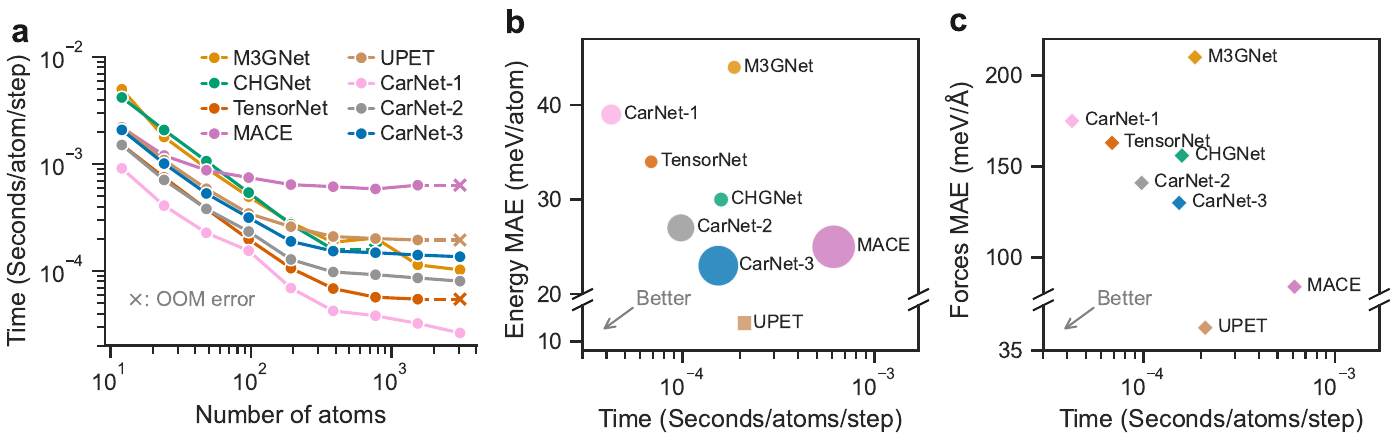}
  \caption{
    \textbf{Accuracy--efficiency trade-off of various universal machine learning interatomic potentials}.
    \textbf{a}.\ Time per molecular dynamics (MD) step normalized by the number of atoms as a function of system size.
    \textbf{b}.\ Test set mean absolute error (MAE) of energy versus time at the system size of 384 atoms.
    \textbf{c}.\ Test set MAE of forces versus time.
    We measure the time by running MD simulations of bulk water systems with different sizes and report the average time per MD step normalized by the number of atoms.
    MD simulations were conducted using the Atomic Simulation Environment (ASE) on a single NVIDIA RTX 5090 GPU; the OOM label indicates out-of-memory errors during the calculations.
    \model-1, \model-2, and \model-3 denote the \model models with 1, 2, and 3 graph neural network (GNN) layers, respectively.
    In panel b, the marker size is proportional to the model size (number of parameters), except for UPET, whose model size is much larger than the others (see \tref{tab:matpes}).
  }
  \label{fig:umlip:matpes}
\end{figure*}

To further demonstrate the capability of \model to model complex systems, we train a universal MLIP for 89 chemical elements using the MatPES r$^2$SCAN dataset~\cite{kaplan2025matpes}.
We use the same data split and train for 100 epochs as in the original MatPES manuscript~\cite{kaplan2025matpes}.
We trained three versions of \model all using $N_u=128$ feature channels and  a maximum natural tensor rank of $L=2$, but with different numbers of GNN layers (1, 2, and 3) to investigate the accuracy--efficiency trade-off.
The three-layer \model took $\sim$34 hours to train on a single NVIDIA RTX 5090 GPU.
Compared with state-of-the-art universal MLIPs trained on this dataset, \model demonstrates highly competitive performance in terms of accuracy.
With only a single GNN layer, it achieves test set MAEs (\tref{tab:matpes}) comparable to those of M3GNet~\cite{chen2022a}, CHGNet~\cite{deng2023chgnet}, and TensorNet~\cite{simeon2023tensornet}.
The three-layer \model attains overall accuracy comparable to MACE~\cite{batatia2025fondation} (which, however, is additionally pretrained on the large OMat24 dataset~\cite{barroso2024open}), with smaller energy and stress errors but larger force errors.
UPET~\cite{mazitov2025petmad} achieves the smallest MAEs among all models.
This advantage is most plausibly attributable to two factors: UPET employs 192.89 million parameters, more than an order of magnitude larger than any other model, and, like MACE, is pretrained on OMat24 (118 million configurations) before being finetuned on MatPES.
Since MACE shares the same pretraining yet performs only on par with \model, UPET's additional edge appears to stem primarily from its much larger model capacity.
Unlike all other models considered here, however, UPET does not guarantee rotationally invariant energy predictions, meaning that the predicted energy of a structure may change with its orientation in space.
Taken together, these comparisons indicate that \model is highly competitive among universal MLIPs.

\model also demonstrates superior efficiency in terms of computational time and memory usage.
We measure the inference time of the models in MD simulations of bulk water systems of different sizes.
The single-layer \model is the fastest model among all, and the three-layer version is faster than MACE and UPET (\fref{fig:umlip:matpes}a).
On the NVIDIA RTX 5090 GPU with 32 GB memory, M3GNet and all \model models can successfully simulate systems up to 3072 atoms (\fref{fig:umlip:matpes}a), while the other models encounter out-of-memory (OOM) errors.
The timing tests were also done for NaCl crystals, and similar results were obtained (see Supplementary Figure~3).
We also investigate the accuracy--efficiency trade-off by plotting the test set MAEs against the inference time at 384 atoms (where all models can run without OOM errors).
For both energy and forces, the \model models together with TensorNet and UPET form the Pareto front (\fref{fig:umlip:matpes}b, c).
In terms of model size (number of parameters), \model is an order of magnitude larger than M3GNet, CHGNet, and TensorNet, and remains slightly smaller than MACE; all are much smaller than UPET (\tref{tab:matpes}).

\section{Discussion}
\label{sec:discussions}

\begin{figure*}[tbh!]
  \centering
  \includegraphics[width=0.95\linewidth]{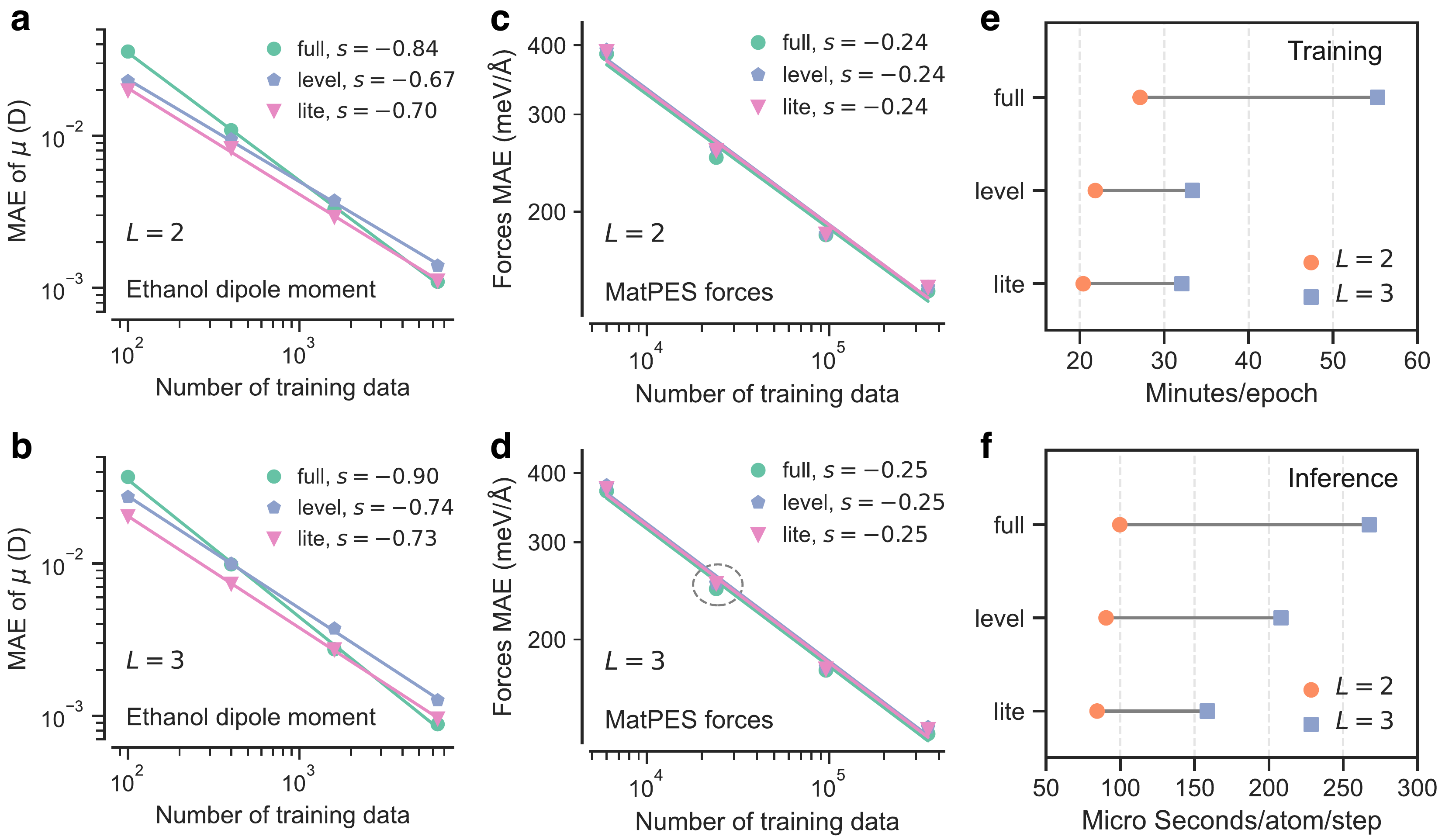}
  \caption{\textbf{Comparison of the tensor product path selection modes.}
    \textbf{a, b}: Mean absolute error (MAE) on learning ethanol dipole moment $\bm\mu$ as a function of the training data size.
    \textbf{c, d}: MAE of forces on the MatPES dataset as a function of the training data size.
    \textbf{e, f}: Training time on the MatPES dataset and inference time using the trained models in molecular dynamics (MD) simulations of bulk water with 384 atoms.
    The slope $s$ is obtained by linearly fitting the learning curve in log-log space.
    $L$ denotes the maximum rank of the natural tensors used in the model.
    Other hyperparameters for training ethanol dipole moment are:
    number of feature channels $N_u=64$, number of layers $T=2$, and cutoff radius $r_\text{cut}= 5$~\AA.
    Other hyperparameters for MatPES training are:
    $N_u=128$, $T=2$, and $r_\text{cut}= 5$~\AA.
  }
  \label{fig:learning:curve}
\end{figure*}

This work develops a framework for atomistic ML by leveraging Cartesian natural tensors to systematically represent high-rank and many-body interactions in atomic structures.
The proposed theory of natural tensor operations enables modeling of a wide variety of physical properties, from scalar quantities (like interatomic potential energy) to tensors of arbitrary rank and symmetry (e.g., the elastic constant tensor)---a capability unmatched by existing Cartesian approaches.  Key design choices enhance computational efficiency while maintaining predictive accuracy, including the implementation of sparsified tensor product paths and the optional incorporation of atomic species dependence in model parameters.

According to the product rule~\cite{coope1970irreducible2}, the rank $l_3$ of the natural tensor $\mathbf{Z}_{l_3}$ resulting from the tensor product of $\mathbf{X}_{l_1}$ and $\mathbf{Y}_{l_2}$  satisfies $|l_1-l_2 | \leq l_3 \leq l_1+l_2$.
Consequently, different pairs $l_1, l_2$ can produce the same $l_3$.
For example, $l_1=1, l_2=1$  and $l_1=1, l_2=2$ can both yield $l_3=1$.
We define a path $p = (l_1, l_2, l_3)$ to denote such a combination.
A central question pertains to the selection of paths that effectively contribute to the computation of atomic and hyper moments using \eref{eq:tp}.
A straightforward approach, analogous to spherical models like NequIP~\cite{batzner2022e}, involves incorporating all possible paths, which ensures maximal expressivity but incurs significant computational cost.
We find that a judicious subset of paths can often achieve superior accuracy and efficiency.

We introduce three path selection modes: `full', `lite', and `level'.
The full mode includes all permissible paths, whereas the lite and level modes use fewer paths by restricting the ones leading to large $l_3$ values.
Detailed descriptions and explicit tabulations of the paths are provided in Supplementary Note~4.
To evaluate their performance, we trained models with these three modes on the ethanol dataset and MatPES dataset, and checked their learning curves.
For ethanol dipole moment, the lite mode yields the lowest MAE when the training data size is small (\fref{fig:learning:curve}a, b)
With a larger learning curve slope $s$ than the other two modes, the full mode's MAE decreases more rapidly with increasing training data, eventually outperforming the lite mode at larger sizes.
For the MatPES dataset, the full mode consistently gives the lowest force MAEs
across all training data sizes (\fref{fig:learning:curve}c, d), although the differences between the modes are less pronounced (e.g.\ 247, 252, and 252 meV/\AA\ for the full, lite, and level modes, respectively, at 24000 training configurations, circled in \fref{fig:learning:curve}d).
Additional results on using varied training settings such as different number of feature channels (Supplementary Figure~4) and on other properties such as energy and stress (Supplementary Figure~5) show the same behavior.
We also compared the efficiency of the three modes, finding that the lite mode is much faster than the full mode in both model training and inference (\fref{fig:learning:curve}d, e).
For example, when using a maximum tensor rank of $L=2$, the lite mode and the full mode take 20.4 and 27.1 minutes per epoch, respectively, to train on the  MatPES dataset;
they take 84 and 99 microseconds per atom per step, respectively, to run an MD simulation of bulk water with 384 atoms.
For $L=3$ with more path reduction, the efficiency gap by the lite mode is even larger.

The different learning-curve trends can be attributed to the interplay between the expressiveness of the path selection modes and the underlying delicacy of the datasets.
The full mode with all tensor product paths possesses a higher degree of expressivity that enables it to resolve the delicate interatomic interactions within the ethanol dataset.
Specifically, it captures the subtle and fine details between molecular conformations along AIMD trajectories that the other two modes can overlook.
As the training data size increases, this higher model capacity allows the full mode to more accurately map these delicate features, manifesting as a steeper learning-curve slope.
Paradoxically, while the MatPES dataset exhibits greater chemical and structural complexity due to its inclusion of 89 elements and diverse crystal structures, it lacks the delicacy like that in the ethanol trajectories.
Consequently, the additional degrees of freedom provided by the full mode remain underutilized on MatPES, leading to a saturation of performance where all three path selection modes result in nearly identical learning curves.
To further validate this, we have conducted additional experiments on the relatively simple LiPS dataset and the more complex elasticity dataset.
Exactly the same trends are observed (Supplementary Figures~6 and~7).
This suggests that the effectiveness of high-rank tensor product paths is governed less by elemental diversity and more by the delicacy of the underlying potential energy surface.
However, dataset complexity, particularly chemical diversity, determines the optimal choice of model parameterization.
For datasets with a few chemical elements like ethanol and LiPS, we observe that making the weights in Eqs.~\eqref{eq:path:mix} and~\eqref{eq:feature:update} dependent
on the atomic number $z_i$ can enhance accuracy.
This, however, does not hold for the MatPES and elasticity datasets with many chemical elements, as it leads to a large number of parameters that cannot be reliably learned from the available data.

Why Cartesian natural tensors?
The field of atomistic ML has traditionally relied on spherical tensor representations and achieved considerable success.
In contrast, the adoption of Cartesian representations lags behind, likely because the theoretical framework of natural tensors is less well-known and not yet fully developed.
Here, we address these issues by extensions to the theory and mathematical formalism of natural tensors, as well as providing practical implementation strategies.
While Cartesian and spherical representations describe the same underlying
spatial reality and are mathematically connected (e.g., the tensor product
$\hat{\mathbf{r}} \otimes \hat{\mathbf{r}} \otimes \dots \otimes \hat{\mathbf{r}}$ can be expanded in terms of
spherical harmonics~\cite{dusson2022atomic}), their mathematical formulations
and computational implementations differ significantly, each offering distinct
advantages and limitations~\cite{zaverkin2024higher}.
In addition, as discussed in \sref{sec:nat:tensor:theory}, natural tensors possess a clear physical interpretability.

There are, however, current limitations that may be further addressed to expand the potential of our approach.
For example, the tensor product path selection scheme is primarily guided by empirical intuition;
although different choices like the lite and level modes are comparable in the number of paths, the lite mode outperforms the level mode in general.
This suggests that a more systematic approach to path selection could lead to further improvements in model performance and efficiency.
Furthermore, other strategies for reducing the computational cost of the tensor product operation have also been proposed in the literature.
For example, So3krates~\cite{frank2022so3krates} achieves efficiency by decoupling invariant features from equivariant coordinates to drive an attention mechanism;
eSCN~\cite{passaro2023reducing} rotates frames into local coordinate systems to achieve a sparse representation;
the Gaunt tensor product method~\cite{luo2024enabling} maps the Clebsch--Gordan coefficients to a 2D Fourier basis, enabling the use of Fast Fourier Transforms to reduce the complexity of full tensor products.
Distinct from these approaches, the high-order pair-reduced neural network~\cite{yang2025high} avoids the tensor product completely by utilizing a hierarchical angular interaction scheme based on direct concatenation and Hadamard products.
Our tensor product path selection scheme adopts a different strategy, which is to sparsify the tensor product paths.
However, while these strategies were developed in the context of Clebsch--Gordan tensor products for spherical models, they may be adapted to the natural tensor product in Cartesian models.
Finally, dedicated GPU kernels have been optimized for spherical tensor operations~\cite{cuEquivariance,openequivariance}, and  similar optimizations may be implemented for natural tensor operations to further improve both training and inference efficiency.

While this work focuses on atomistic ML, the proposed Cartesian natural tensor framework is broadly applicable to other domains.
Atomistic ML can be viewed as a specific instance of point cloud learning, where each point (atom) possesses attributes such as atomic number and spatial coordinates.
Consequently, the natural tensor formalism and the \model architecture can be extended to a variety of other point cloud tasks, such as shape learning in 3D medical images~\cite{chen2021shape,zhang2025hierarchical} and object recognition and segmentation in LiDAR data for autonomous driving applications~\cite{qi2017CVPR,shi2020CVPR}.

\section*{Methods}
\label{sec:methods}

\subsection*{Dataset}

The LiPS dataset~\cite{batzner2022e} consists of 250001 structures of lithium phosphorus sulfide (\ce{Li_{6.75}P3S11}) solid-state electrolyte, generated from an AIMD trajectory. Each structure consists of 27 Li atoms, 12 P atoms, and 44 S atoms.
Random subsets of 1000, 1000, and 5000 structures were selected for training, validation, and test, respectively.

The water dataset~\cite{cheng2019ab} consists of 1593 water configurations of 192 atoms each, obtained from AIMD simulations at 300~K.
It was randomly divided into training, validation, and test sets with a split ratio of 90:5:5.

The ethanol dataset~\cite{gastegger2021machine} includes 10000 molecular structures, with five target properties (energy, forces, dipole moment, polarizability, and nuclear shielding), all computed in vacuum using DFT.
It was randomly divided into training, validation, and test sets with a split ratio of 8:1:1.

The MatPES r$^2$SCAN dataset~\cite{kaplan2025matpes} consists of 387897 DFT calculations of 89 elements in various crystal structures.
We use the same training, validation, and test splits as in the original work~\cite{kaplan2025matpes}, which are 80\%, 10\%, and 10\% of the total dataset, respectively.

The elasticity dataset~\cite{wen2025cartesian} contains elastic constant tensors for inorganic crystals from the DFT data in the Materials Project~\cite{horton2025accelerated}.
It includes 10276 elastic constant tensors, and we use the same training, validation, and test splits as in the original work~\cite{wen2024an}, which are 80\%, 10\%, and 10\% of the total dataset, respectively.

\subsection*{Model Architecture}
\label{sec:model}

\textbf{Atomic embedding}
An atomic structure is represented as a graph $G = (V, E)$, where the nodes $V$ correspond to atoms, and the edges $E$ connect pairs of atoms within a cutoff radius $r_\text{cut}$.
A node $i$ (atom) is characterized by three properties: the atomic coordinates $\mathbf{r}^i$, atomic number $z_i$, and atom features $\mathbf{h}^i$.
For each edge $(i,j)$ we define the relative position vector as: $\mathbf{r}^{ij} = \mathbf{r}^j - \mathbf{r}^i$, which encodes the spatial relationship between atom $i$ and its neighbor $j$.

The atom features $\mathbf{h}^i$ consist of a set of natural tensors, indexed by $u$ and $l$.
In their tensorial form, these features are represented as $\mathbf{h}^i_{ul}$ where: $l$ denotes the tensor rank and $u$ labels the feature channels.
The initial features of each atom are derived by embedding its atomic number:
\begin{equation}
  \mathbf{h}_{u0}^i =  W_{u z^i} ,
\end{equation}
where $W_{uz^i}$ is a learnable embedding matrix.
These initial atom features are scalar-valued and constitute natural tensors of rank $l=0$.

\textbf{Angular part}
The unit edge vector $\hat{\mathbf{r}}^{ij} = \mathbf{r}^{ij} / r^{ij}$ with $r^{ij} = \|\mathbf{r}^{ij}\|$ encodes the directional information between atoms $i$ and $j$.
Atom indices $i$ and $j$ in $\mathbf{r}^{ij}$ are omitted for simplicity.
To capture angular interactions of various orders, we construct the polyadic tensor
$
  \mathbf{U}_l = \hat{\mathbf{r}} \otimes \hat{\mathbf{r}} \otimes \dots \otimes \hat{\mathbf{r} }
$
where the tensor product is taken $l$ times, producing a rank-$l$ tensor.
Each $\mathbf{U}_l$ can be decomposed into its natural tensor components, resulting in the set:
\begin{equation}
  \mathbf{X}_0 = 1 ,
  \
  \mathbf{X}_1 =
  \begin{bmatrix}
    \hat r_x \\
    \hat r_y \\
    \hat r_z \\
  \end{bmatrix},
  \
  \mathbf{X}_2 =
  \begin{bmatrix}
    \hat r_x^2-\sigma_0 & \hat r_x \hat r_y   & \hat r_x \hat r_z   \\
    \hat r_y\hat r_x    & \hat r_y^2-\sigma_0 & \hat r_y \hat r_z   \\
    \hat r_z\hat r_x    & \hat r_z \hat r_y   & \hat r_z^2-\sigma_0
  \end{bmatrix}
\end{equation}
and so forth, where $\hat r_x$, $\hat r_y$ and $\hat r_z$ are the Cartesian components of $\hat {\mathbf{r}}$, and $\sigma_0 = (\hat r_x^2 + \hat r_y^2 + \hat r_z^2)/3$, which is one-third the trace of $\mathbf{U}_2$---ensuring $\mathbf{X}_2$ is traceless (more explicitly, $\mathbf{X}_l=\mathbf{X}^{ij}_l$).

\textbf{Radial basis}
The interatomic distance (edge length) $r^{ij}$ is expanded in a set of radial basis functions $B_u$ indexed by channel $u$.
Basis functions are constructed as linear combinations of Chebyshev polynomials of the first kind $Q_\beta$,
\begin{equation} \label{eq:radial:basis}
  B_u(r) = \sum_{\beta=0}^{N_\beta} W_{u\beta} Q_\beta \left(\frac{r}{r_\text{cut}} \right) f_c \left( \frac{r}{r_\text{cut}} \right),
\end{equation}
where $N_\beta$ is the maximum degree of the Chebyshev polynomials and $W_{u\beta}$ are learnable weights.
The radial expansion is similar to those used in MTP~\cite{novikov2020mlip} and CAMP~\cite{wen2025cartesian},
but here the weights $W_{u\beta}$ are shared across different atomic species, significantly reducing the total number of learnable parameters; this is advantageous for datasets with many chemical elements.

\textbf{Atomic moment}
Using the atom features, angular information, and radial basis, we construct an atomic moment that encodes the local environment of each atom:
\begin{equation} \label{eq:atomic:moment}
  \mathbf{M}^i_{ul_3, p}
  = \frac{1}{ \sqrt{|\mathcal{N}|}}
  \sum_{j \in \mathcal{N}_i} R_{ul_3l_1l_2} \mathbf{h}^j_{ul_1} \hat \otimes \mathbf{X}^{ij}_{l_2},
\end{equation}
where $\mathcal{N}_i$ is the set of neighboring atoms within a distance of $r_\text{cut}$ of atom $i$,
$|\mathcal{N}|$ is the average number of neighbors per atom in the training set,
$R_{ul_3l_1l_2}$ is a learnable radial function,
and
$\hat\otimes$ denotes the natural tensor product (as defined in \sref{sec:nat:tensor:theory}).
The radial function $R_{ul_3l_1l_2}$ is obtained by passing the radial basis $B_u$ through a multilayer perceptron (MLP): $R_{ul_3l_1l_2} = \text{MLP}
  (B_u)$ using two hidden layers with the SiLU nonlinearity~\cite{elfwing2017silu}.
Different MLPs are used for each combination of indices $(l_3, l_2, l_1)$.
Similar to spherical tensor products, the product of natural tensors of ranks $l_1$ and $l_2$ can produce a natural tensor of rank $l_3$.
The triplet $p = (l_1, l_2, l_3)$ is called a path and determines the tensorial combination in \eref{eq:atomic:moment}.

Atomic moment tensors of the same rank but originating from different tensorial paths are linearly combined as follows:
\begin{equation} \label{eq:path:mix}
  \mathbf{M}_{ul}^i = \sum_{u'p} W_{uu'l,p}^{z_i}\mathbf{M}_{u'l,p}^i,
\end{equation}
where $W_{uu'l,p}^{z_i}$ are trainable weights.
To reduce the number of parameters, these weights are factored as $W_{uu'l,p}^{z_i} = W_{uu'l}^{z_i} W_p$.

\textbf{Hyper moment}
From the atomic moments, we construct the hyper moment,
\begin{equation} \label{eq:hyper:moment}
  \mathbf{H}_{ul}^i = \mathbf{M}_{ul_1}^i \hat\otimes \mathbf{M}_{ul_2}^i \hat\otimes \cdots \hat\otimes \mathbf{M}_{ul_v}^i \quad (\text{$v$ of $\mathbf{M}$}).
\end{equation}
where $\hat{\otimes}$ denotes the natural tensor product and $v$ is the number of atomic moments being combined.
An atomic moment encodes two-body interactions between an atom and its neighbors.
By taking tensor product of atomic moments with themselves, higher-order interactions are incorporated: three-body, four-body, and beyond.
Specifically, a hyper moment of degree $v$ captures interactions up to body order $v+1$.
The hyper moment thus provides a systematic and complete description of the local atomic environment~\cite{shapeev2016moment,dusson2022atomic}, which is essential for constructing systematically improvable interatomic potentials.
This approach is analogous to the B-basis used in ACE~\cite{drautz2019atomic} and MACE~\cite{batatia2022mace}.
Similar to atomic moments, multiple tensor paths can generate hyper moments of the same rank $l$ that may be linearly combined.
An efficient algorithm for evaluating \eref{eq:hyper:moment} iteratively is provided in Supplementary Note~5.

\textbf{Feature update}
Atom features are then updated using a residual connection~\cite{he2015deep} by combining the hyper moments with atom features of the previous layer:
\begin{equation} \label{eq:feature:update}
  \mathbf{h}^{i,t}_{ul} =  \mathbf{H}_{ul}^i + \sum_{u'} W_{uu'l}^{z_i} \mathbf{h}^{i,t-1}_{u'l},
\end{equation}
where $t$ is the layer index.

\textbf{Output Construction}
The feature update process is performed for $N_\text{layer}$ layers, producing a sequence of atom features:
$\mathbf{h}_{ul}^{i,1}$,
$\mathbf{h}_{ul}^{i,2}$,
$\dots$,
$\mathbf{h}_{ul}^{i,N_\text{layer}}$.
Depending on the modeling target, either the features from all layers or a subset are used to construct the final output.
Empirically, two or three layers are sufficient for interatomic potentials, atomic tensors, and structure tensors.

For interatomic potentials, the atomic energy $E^i$ is derived from  $l=0$ scalar atom features $\mathbf{h}^{i,t}_{u0}$ across all layers
\begin{equation}
  E^i = \sum_{t=1}^{N_\text{layer}} V(\mathbf{h}^{i,t}_{u0}),
\end{equation}
in which $V$ is implemented as an MLP using two hidden layers with the SiLU nonlinearity for the last layer where $t=N_\text{layer}$,
and a linear function,
$V(\mathbf{h}^{i,t}_{u0}) = \sum_u W^t_u \mathbf{h}^{i,t}_{u0}$, for earlier layers ($t<N_\text{layer}$).
The total potential energy is the sum over all atoms:
\begin{equation} \label{eq:ip:output}
  E = \sum_i \sigma E^i + \mu_{z_i},
\end{equation}
where $\mu_{z_i}$ is the atomic energy for species $z_i$ (obtained directly from DFT calculations or computed via a linear fit to the total energies of the training set),
and $\sigma$ is the root mean square of the atomic forces computed on the training set~\cite{musaelian2023learning,batatia2025fondation}.
Forces are obtained via the negative gradient:
$\mathbf{F}_i = - \frac{\partial E}{\partial \mathbf{r}_i}. $

For atomic tensors, the atom features
$\mathbf{h}_{ul}^{i,1}$,
$\mathbf{h}_{ul}^{i,2}$,
$\dots$,
$\mathbf{h}_{ul}^{i,N_\text{layer}}$
from all layers are linearly combined along the channel dimension $u$ to produce a rank-$l$ natural tensor for each atom:
\begin{equation} \label{eq:atomic:natural:tensor}
  \mathbf{v}_{l}^i  = \sum_{t,u} W_{u}^t \mathbf{h}_{ul}^{i,t}.
\end{equation}
Relevant natural tensors are then selected to reconstruct physical tensors.

For example, the rank-0 $\mathbf{v}_0^i$, rank-1 $\mathbf{v}_1^i$ and rank-2 $\mathbf{v}_2^i$ natural tensors are used to reconstruct the rank-2 nuclear shielding tensor $\bm\sigma$ for each atom $i$.
For the shielding tensor, there is only one unique independent mapping tensor for each rank, i.e., $N_g=1$ for all $m$ in \eref{eq:reconstruct}.
Therefore, the reconstruction of the shielding tensor is given by:
\begin{equation}
  \bm \sigma^i = \sum_{m=0}^2 \mathbf{Q}_{m+2} \odot^m \mathbf{X}_m
\end{equation}
where $\mathbf{X}_0 = \mathbf{v}_0^i$, $\mathbf{X}_1 = \mathbf{v}_1^i$, and $\mathbf{X}_2 = \mathbf{v}_2^i$ are the natural tensors of rank 0, 1, and 2, respectively.

For structural tensors, natural tensors of different ranks $l$ are first generated for each atom according to \eref{eq:atomic:natural:tensor}.
They are then aggregated across all atoms by taking the average or sum, depending on whether the target physical tensor is intensive or extensive.
Then relevant natural tensors are selected to reconstruct the physical tensor.

For extensive quantities (e.g., the dipole moment $\mathbf{u}$ and polarizability $\bm\alpha$), the tensors are summed:
\begin{equation}
  \mathbf{X}_l = \sum_i \mathbf{v}_l^i.
\end{equation}

According to the decomposition and recombination spectrum (given in Supplementary Table~3),
the dipole moment $\mathbf{u}$ only needs the rank-1 natural tensor $\mathbf{X}_1$ for its reconstruction.
It can be obtained via \eref{eq:reconstruct} as:
\begin{equation}
  \mathbf{u} = \mathbf{Q}_{m+1} \odot^m \mathbf{X}_m,
\end{equation}
where $m=1$.
The polarizability $\bm\alpha$ is a symmetric rank-2 tensor, which can be reconstructed from the rank-0 and rank-2 natural tensors ($\mathbf{X}_0$ and $\mathbf{X}_2$) as:
\begin{equation}
  \bm\alpha = \sum_{m\in\{0,2\}} \mathbf{Q}_{m+2} \odot^m \mathbf{X}_m.
\end{equation}

For the intensive elastic constant tensor $\mathbf{C}$, the average is taken:
\begin{equation} \label{eq:intensive:structure:tensor}
  \mathbf{X}_l = \frac{1}{N}\sum_i \mathbf{v}_l^i,
\end{equation}
where $N$ is the total number of atoms in the structure.
The rank-4 elastic constant tensor $\mathbf{C}$ is more complex: it decomposes into two rank-0 natural tensors ($\mathbf{X}_0^1$ and $\mathbf{X}_0^2$), two rank-2 natural tensors ($\mathbf{X}_2^1$ and $\mathbf{X}_2^2$), and a rank-4 natural tensor $\mathbf{X}_4$ (see Supplementary Table~3).
The two tensors $\mathbf{X}_0^1$ and $\mathbf{X}_0^2$ of the same rank are obtained by employing separate $W_u^t$ in \eref{eq:atomic:natural:tensor}; similarly for $\mathbf{X}_2^1$ and $\mathbf{X}_2^2$.
Then the elastic constant tensor is reconstructed via \eref{eq:reconstruct} as:
\begin{equation}
  \mathbf{C} = \sum_{m \in \{0,2,4\}} \sum_{g=1}^{N_g(m)} \mathbf{Q}^g_{m+4} \odot^m \mathbf{X}^g_m,
\end{equation}
where $N_g(0)=2$, $N_g(2)=2$, and $N_g(4)=1$.

\subsection*{Normalization}

\textbf{Internal normalization.}
Properly normalizing the internal features of a neural network is critical for ensuring stable and efficient training.
In this work, we adopt the default uniform initialization scheme in PyTorch~\cite{paszke2019pytorch} for all learnable weights, but we update the initialization bounds so that all network components yield outputs with approximately zero mean and unit variance.
Furthermore, when summing over the features of neighboring atoms to compute the atomic moment in \eref{eq:atomic:moment}, we normalize the sum by dividing by the square root of the average number of neighbors, $\sqrt{|\mathcal{N}|}$, in the training set.
This normalization assumes that contributions from different terms are uncorrelated, such that their variances are additive~\cite{musaelian2023learning}.

\textbf{Target normalization.}
Normalizing learning targets is equally crucial for model performance~\cite{musaelian2023learning}.
This normalization process is equivalent to the choice of scale and shift parameters for the model's output.
For interatomic potentials, we use the root mean square of the atomic forces $\sigma$ as the scale parameter and the atomic energy $\mu_{z_i}$ as the shift parameter, as described in \eref{eq:ip:output}.

For tensorial properties, we perform normalization in the natural tensor space rather than the Cartesian physical tensor space, using rank-dependent scale and shift parameters.
For scalar natural tensors ($l=0$), the shift $\mu_l$ and scale $\sigma_l$ are set to the mean and standard deviation of the target scalar values across the training set.
For higher-rank natural tensors ($l>0$), the shift is set to zero ($\mu_l=0$) to maintain equivariance, while the scale $\sigma_l$ is set to the root mean square of the Frobenius norm of the target natural tensors of the same rank.
Applying these parameters to the atomic natural tensors, \eref{eq:atomic:natural:tensor} becomes:
$
  \mathbf{v}_{l}^i = \sigma_l \left( \sum_{t,u} W_{u}^t \mathbf{h}_{ul}^{i,t} \right) + \mu_l
$.
We note that although this normalization is conducted at the atomic level, the formulation remains valid for structural targets, which are simply linear combinations of atomic values. For example, under this normalization, \eref{eq:intensive:structure:tensor} becomes:
$
  \mathbf{v}_l = \frac{1}{N}\sum_i \left(\sigma_l \mathbf{v}_{l}^i + \mu_l \right) = \sigma_l \left(\frac{1}{N}\sum_i \mathbf{v}_{l}^i\right) + \mu_l,
$
which is identical to normalizing the structural tensor directly.

\subsection*{Model Training}

Interatomic potentials are trained by minimizing a loss function of energy and forces (and stress if available).
For a given atomic structure, the loss is
\begin{equation}
  l(\theta) =
  w_\mathcal{E} \left( \frac{ \mathcal{E} - \hat{\mathcal{E}}}{N} \right)^2
  + w_\text{F} \frac{\sum_{i=1}^N \| \mathbf{F}_i - \hat{\mathbf{F}_i} \|^2}{3N}
  + w_\text{S}  \| \mathbf{S} - \hat{\mathbf{S}} \|^2,
\end{equation}
where $N$ is the number of atoms, $\mathcal{E}$, $\mathbf{F}_i$, and $\mathbf{S}$ are the model predicted energy, forces, and stress,
$\hat{\mathcal{E}}$, $\hat{\mathbf{F}_i}$, and $\hat{\mathbf{S}}$ are the corresponding reference values,
and $w_\mathcal{E}$,  $w_\text{F}$, $w_\text{S}$ are the  energy, forces and stress weighting factors.

For atomic tensor properties (e.g., nuclear shielding tensor), the loss is computed as
\begin{equation}
  l(\theta) = \frac{1}{N} \sum_{i=1}^N \| \bm \sigma^i - \hat{\bm \sigma}^i \|^2,
\end{equation}
where $\bm \sigma^i$ is the predicted Cartesian rank-2 nuclear shielding tensor for atom $i$, and $\hat{\bm\sigma}^i$ is its reference value.
For structural tensor properties (e.g., dipole moment, polarizability, and elastic constant tensors), the loss per structure is
\begin{equation}
  l(\theta) =  \| \mathbf{T} - \hat{\mathbf{T}} \|^2,
\end{equation}
where $\mathbf{T}$ and $\hat{\mathbf{T}}$ are the predicted and reference tensors.
The total loss to minimize at each optimization step is the mean of the losses of a minibatch of configurations.

For the elastic constant tensor dataset and the MatPES dataset, we also used the Huber loss~\cite{huber1964robust} for training, which is less sensitive to outliers than the mean squared error loss.
It is defined as
\begin{equation}
  l_\delta =
  \begin{cases}
    \frac{1}{2} \|  y - \hat y \|^2,                & \text{for } \|  y - \hat y \| < \delta \\
    \delta \| y - \hat y \| - \frac{1}{2}\delta^2 , & \text{otherwise}
  \end{cases},
\end{equation}
where $y$ and $\hat y$ denote the predicted and reference values, respectively (e.g., $\mathbf{T}$ and $\hat{\mathbf{T}}$ for elastic constant tensor),
$\delta$ is a hyperparameter that determines the threshold for switching between the quadratic and linear loss regimes.

The models are implemented in PyTorch~\cite{paszke2019pytorch} and trained using PyTorch Lightning~\cite{lightning}.
Optimization is performed with the AdamW optimizer~\cite{kingma2014adam}.
A cosine annealing learning rate schedule is employed, starting with an initial learning rate depending on the specific task.
During model performance evaluation, an exponential moving average of the model weights is maintained, with decay rates of 0.999 for interatomic potential models and 0.99 for tensor property models.
Hyperparameters are selected based on model performance on the validation set, and all results correspond to the test set.

Detailed model training hyperparameters for all tasks are provided in Supplementary Table~1.

\subsection*{Molecular Dynamics}

MD simulations are performed in the NVT ensemble using the Nos\'e--Hoover thermostat.
For LiPS, the simulation employs a cell size identical to that used in the training data, consisting of 83 atoms.
The system is maintained at 520~K with a timestep of 1~fs and a Nos\'e--Hoover chain damping time of 20~fs.
The total simulation duration is 300~ps.
The thermostat is implemented with the ASE \texttt{NoseHooverChain}~\cite{larsen2017atomic}, using a single chain.
Simulations for water are similarly conducted
using a cell with 512 molecules at temperatures from 300--2000~K.

The diffusion coefficient $D$ is calculated from the mean squared displacement (MSD) using the Einstein relation~\cite{einstein1905uber}:
\begin{equation}
  D = \lim_{t\rightarrow\infty} \frac{\left \langle  \frac{1}{N} \sum_i^N \left |\mathbf{r}_i(t) - \mathbf{r}_i(0) \right | ^{2} \right \rangle} {2nt},
\end{equation}
where $N$ is the number of diffusing atoms (lithium for LiPS, oxygen for water),
$\mathbf{r}_i(t)$ is the position of atom $i$ at time $t$, $\langle \cdot \rangle$ denotes an ensemble average over multiple time origins or trajectories, and $n=3$ indicates diffusion in three dimensions.
In practice, the diffusion coefficient is computed from the slope of a linear fit to the MSD versus $2nt$, as implemented in ASE~\cite{larsen2017atomic}.
The data from the initial 10~ps is discarded; the remaining data is used in the fitting (\fref{fig:bulk:result}c, f).
This ensures an accurate estimation of the diffusion coefficient from the long-time linear regime of the MSD.

\section*{Data availability}
All datasets used in this work are publicly available.
The LiPS dataset: \url{https://archive.materialscloud.org/record/2022.45},
the water dataset: \url{https://doi.org/10.1073/pnas.1815117116},
the ethanol dataset: \url{http://quantum-machine.org},
the elasticity dataset: \url{https://doi.org/10.5281/zenodo.8190849},
and the MatPES dataset (v2025.1) is at \url{https://matpes.ai}.

The \model models generated in this study have been deposited in a GitHub repository at \url{https://github.com/wengroup/carnet_run}.
The MACE and UPET models trained on the OMAT dataset and finetuned on the MatPES dataset are also publicly available.
The `MACE-matpes-r2scan-omat-ft.model' checkpoint is at: \url{https://github.com/ACEsuit/mace-foundations/releases/tag/mace_matpes_0},
and the `pet-omatpes-l-v0.1.0.ckpt' checkpoint is at: \url{https://huggingface.co/lab-cosmo/upet/tree/main/models}.

\section*{Code availability}
The code for natural tensor operations is at \url{https://github.com/wengroup/natt}.
The \model source code is at \url{https://github.com/wengroup/carnet}; the version used to generate the results reported in this work, commit \verb|186071c81b49cd1d254c751ed4a4f6b8f1b85df9|, is archived at Zenodo~\cite{chen2026carnet}.
Scripts for training models, running MD simulations, and analyzing the results are at \url{https://github.com/wengroup/carnet_run}.

\section*{Author contributions}

Q.C.: model training, data analysis, and writing - review;
A.S.L.S.P: data analysis, visualization, and writing - review;
B.W.: data analysis and writing - review;
D.J.S.: writing - review;
M.W.: project conceptualization, model development, data analysis, visualization, writing - original draft, writing - review, and supervision.

\section*{Competing Interest}
The authors declare no competing interests.

\section*{Acknowledgements}
This work is supported by the Center for HPC at the University of Electronic Science and Technology of China.
It also uses computational resources provided by the Hefei Advanced Computing Center.

\def\bibsection{\section*{\refname}}

%

\end{document}


\title{\Large Supplementary Information:\\
   Atomistic Machine Learning with Irreducible Cartesian Natural Tensors}

\author{Qun Chen}
\affiliation{Institute of Fundamental and Frontier Sciences, University of Electronic Science and Technology of China, Chengdu, China}

\author{A. S. L. Subrahmanyam Pattamatta}
\affiliation{Department of Mechanical Engineering, The University of Hong Kong, Hong Kong SAR, China}

\author{Boyu Wang}
\affiliation{Institute of Fundamental and Frontier Sciences, University of Electronic Science and Technology of China, Chengdu, China}

\author{David J. Srolovitz}
\affiliation{Department of Mechanical Engineering, The University of Hong Kong, Hong Kong SAR, China}

\author{Mingjian Wen}
\email{mjwen@uestc.edu.cn}
\affiliation{Institute of Fundamental and Frontier Sciences, University of Electronic Science and Technology of China, Chengdu, China}

\maketitle

%

\clearpage

\begin{figure}[H]
   \centering
   \includegraphics[width=0.6\columnwidth]{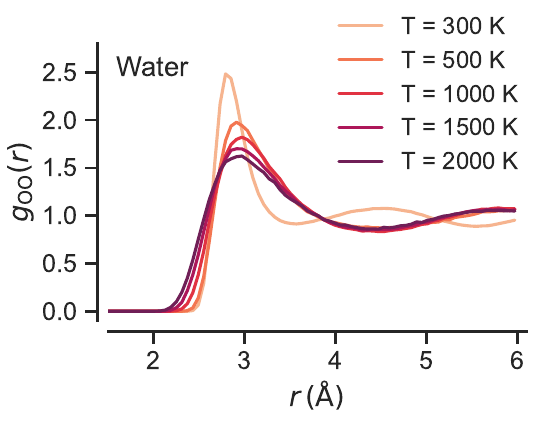}
   \caption[Water RDF at different temperatures]{Radial distribution function (RDF) of water computed from the MD simulations using the \model model at different temperatures.}
\end{figure}

\begin{figure}[H]
   \centering
   \includegraphics[width=0.55\columnwidth]{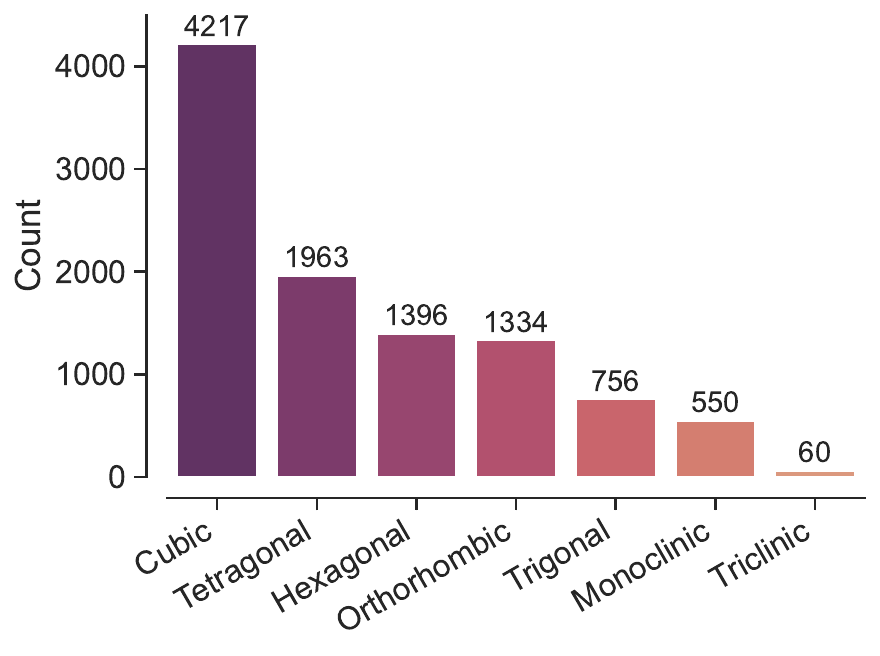}
   \caption[Elastic tensor dataset by crystal system]{Histogram of the elastic tensor dataset categorized by crystal system, showing the number of structures (y-axis) for each crystal symmetry (x-axis).}
\end{figure}

\begin{figure}[H]
   \centering
   \includegraphics[width=0.5\columnwidth]{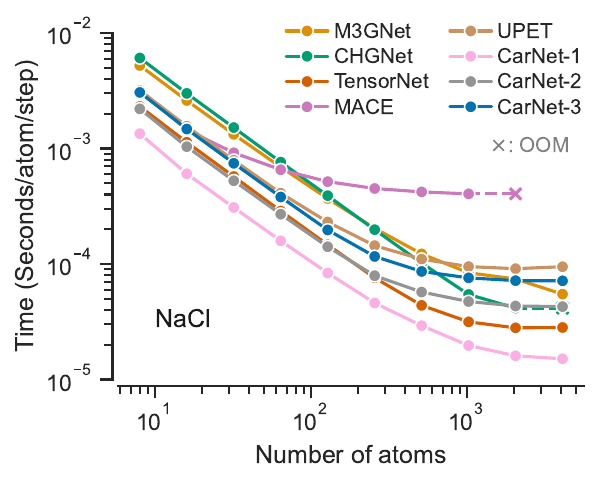}
   \caption[Inference time for NaCl systems]{Inference time of various universal MLIPs for NaCl systems with different model size.
      Out-of-memory (OOM) occurs for some models when the system size is large.
   }
\end{figure}

\begin{figure}[H]
   \centering
   \includegraphics[width=.95\columnwidth]{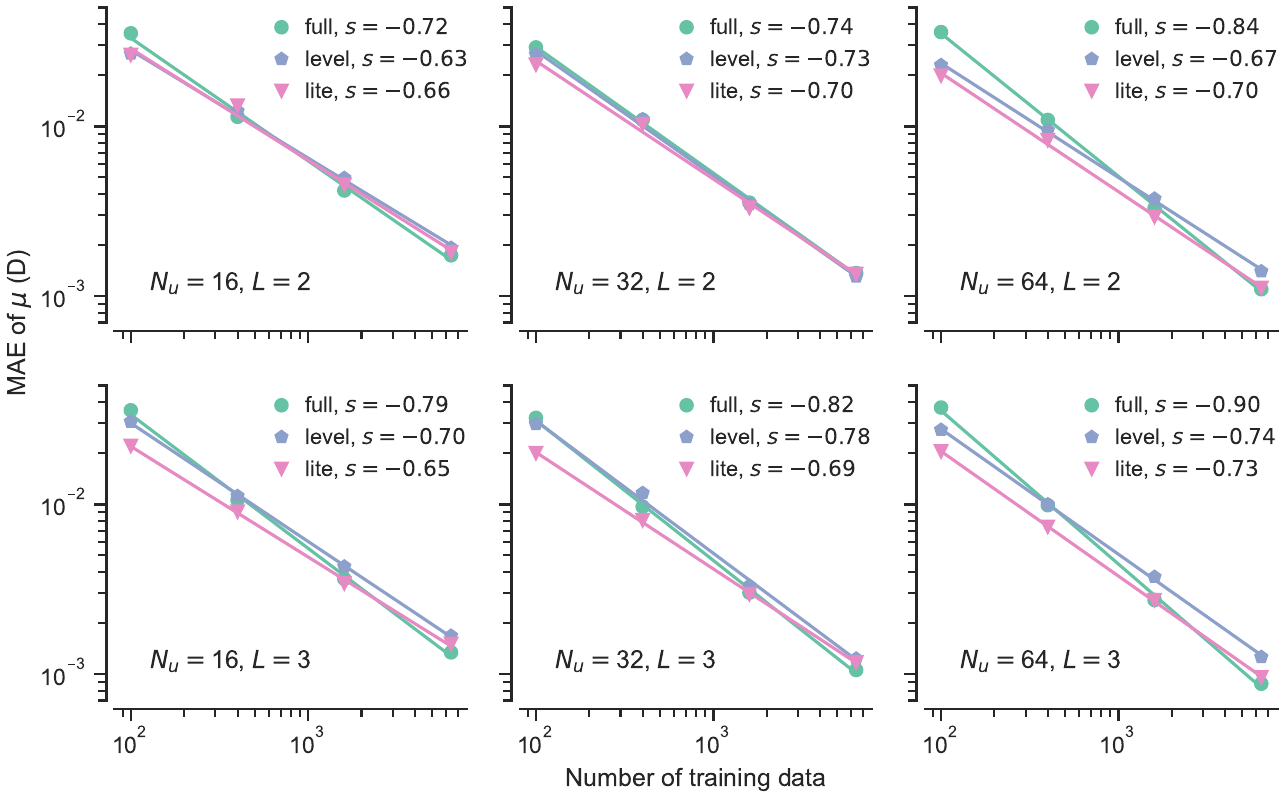}
   \caption[Ethanol dipole moment learning curve]{Learning curve of \model evaluated on ethanol dipole moment $\mu$.
      The MAE is computed on the test set and plotted against the number of data samples used to train the model.
      $N_u$ is the number of feature channels, and $L$ is the maximum allowed rank of the natural tensors used in the model.
      Other model training hyperparameters are: number of layers $T=2$, correlation degree $v = 3$, and cutoff radius $r_\text{cut}= 5$~\AA.
   }
\end{figure}

\begin{figure}[H]
   \centering
   \includegraphics[width=.78\columnwidth]{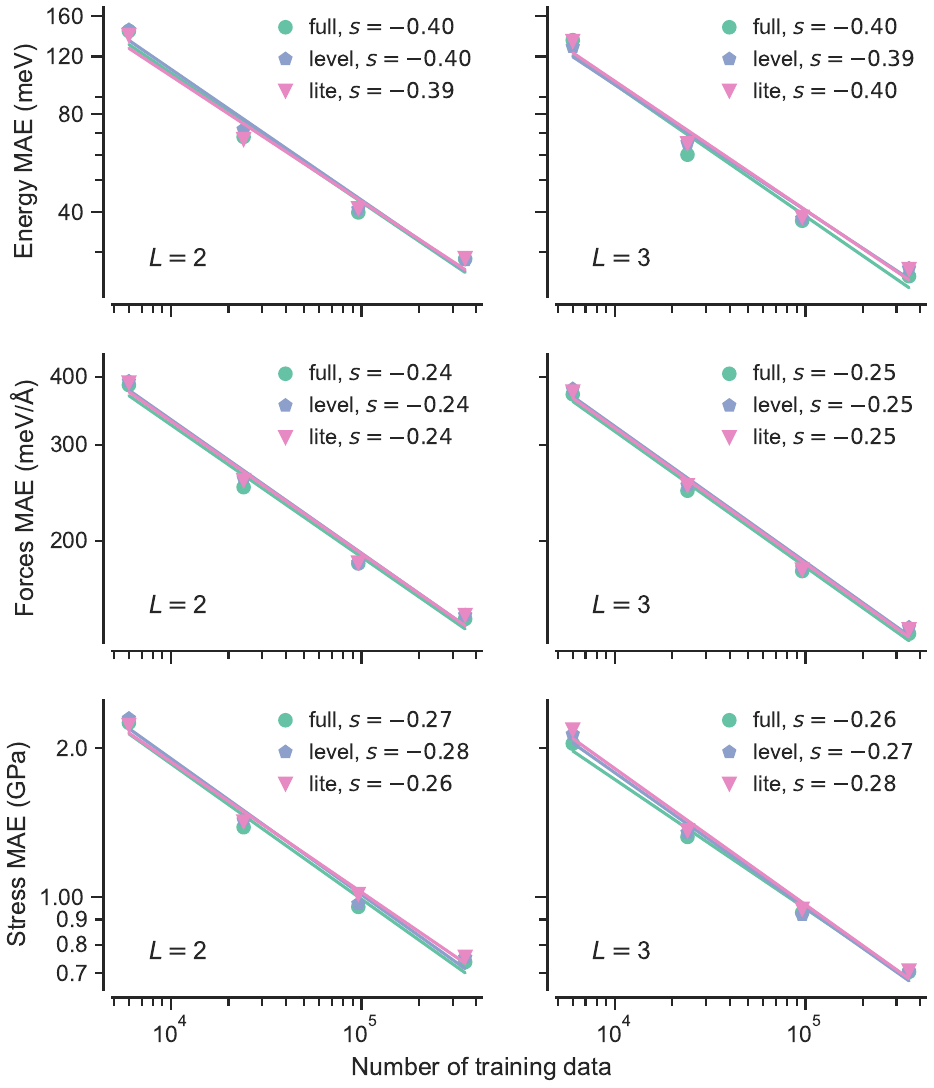}
   \caption[MatPES learning curve]{Learning curve of \model evaluated on the MatPES dataset.
      The MAE is computed on the test set and plotted against the number of data samples used to train the model.
      $L$ denotes the maximum allowed rank of the natural tensors used in the model.
      Other model training hyperparameters are:
      number of feature channels $N_u=128$, number of layers $T=2$, correlation degree $v = 3$, and cutoff radius $r_\text{cut}= 5$~\AA.
   }
\end{figure}

\begin{figure}[H]
   \centering
   \includegraphics[width=0.8\columnwidth]{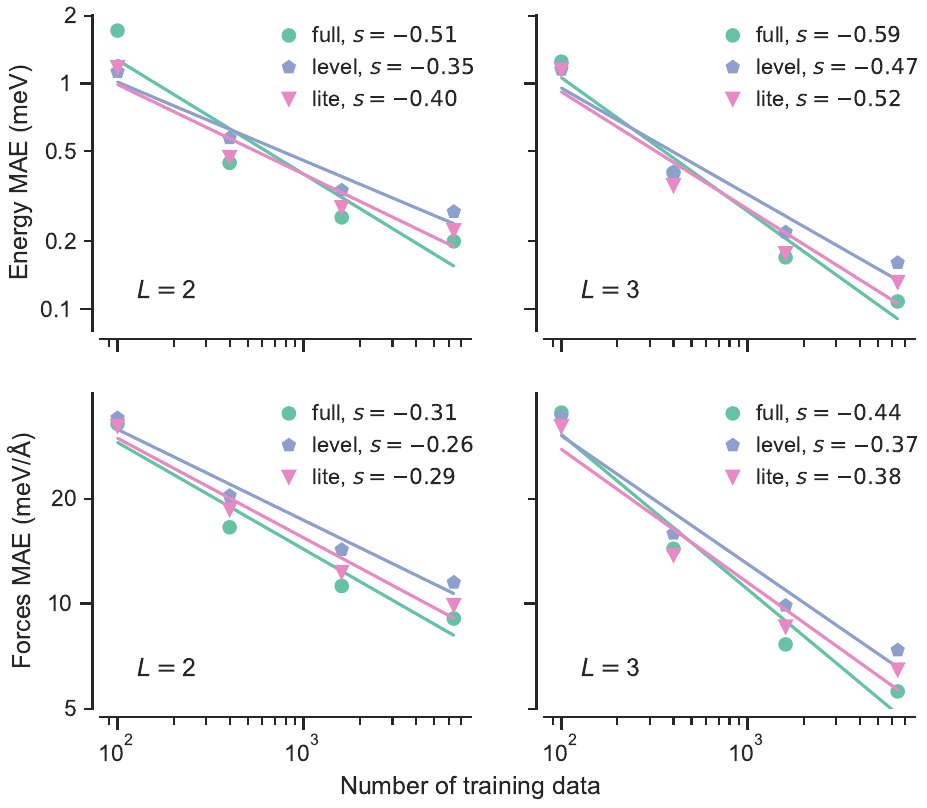}
   \caption[LiPS learning curve]{Learning curve of \model evaluated on the LiPS dataset.
      The MAE is computed on the test set and plotted against the number of data samples used to train the model.
      $L$ denotes the maximum allowed rank of the natural tensors used in the model.
      Other model training hyperparameters are:
      number of feature channels $N_u=32$, number of layers $T=2$, correlation degree $v = 3$, and cutoff radius $r_\text{cut}= 5$~\AA.
   }
\end{figure}

\begin{figure}[H]
   \centering
   \includegraphics[width=0.5\columnwidth]{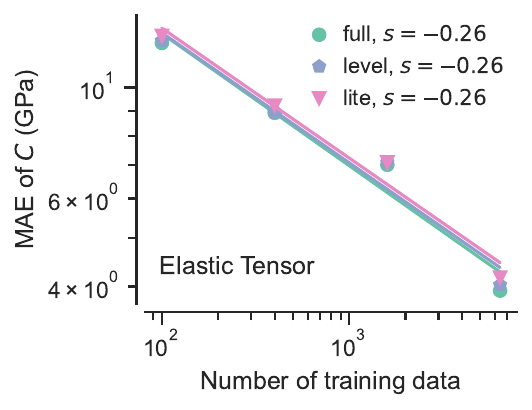}
   \caption[Elastic tensor learning curve]{Learning curve of \model evaluated on the elastic tensor dataset.
      The MAE is computed on $6\times6$ Voigt components of the elastic tensor on the test set.
      Hyperparameters used in model training are: maximum rank of natural tensors $L=4$, number of feature channels $N_u=48$, number of layers $T=2$, correlation degree $v = 3$, and cutoff radius $r_\text{cut}= 5$~\AA.
   }
\end{figure}

\setlength{\tabcolsep}{4pt}
\begin{table}[tbh!]
   \centering
   \caption[Hyperparameters and training details]{Hyperparameters and training Details.
      `Weight mode' determines the type of weights used in Eqs.~12 and 14 in the main text.
      Mode 1 (default): The weights depend on the atomic number $ z $, denoted as $ W_{uu'l,p}^{z_i} $ and $ W_{uu'l}^{z_i} $.
      Mode 2: The weights do not depend on the atomic number $ z $, becoming $ W_{uu'l,p} $ and $ W_{uu'l} $
   }
   \footnotesize
   \begin{tabular}{lccccccc}
      \hline
      Dataset                       & LiPS                                                                                                                                                                                              & Water            & Ethanol                  & Ethanol                          & Ethanol                                                                & Elasticity                                                           & MatPES                        \\
                                    &                                                                                                                                                                                                   &                  & ($\mathcal{E}$, $\mathbf{F}$) & ($\bm\mu, \bm\alpha, \bm\sigma$) & (multitask)                                                            &                                                                                                      \\
      \hline
      TP mode                       & full                                                                                                                                                                                              & lite             & full                     & full                             & full                                                                   & lite                                                                 & lite                          \\
      Weight mode                   & 1                                                                                                                                                                                                 & 1                & 1                        & 1                                & 1                                                                      & 2                                                                    & 2                             \\
      Max rank $L$                  & 3                                                                                                                                                                                                 & 2                & 3                        & 3                                & 3                                                                      & 4                                                                    & 2                             \\
      Corr. degree $v$              & 3                                                                                                                                                                                                 & 2                & 3                        & 3                                & 3                                                                      & 3                                                                    & 3                             \\
      \# layers $T$                 & 2/3                                                                                                                                                                                               & 2/3              & 2/3                      & 3                                & 3                                                                      & 3                                                                    & 2/3                           \\
      \# channels $N_u$             & 48                                                                                                                                                                                                & 48               & 64                       & 64                               & 128                                                                    & 64                                                                   & 128                           \\
      \# Chebyshev basis            & 8                                                                                                                                                                                                 & 8                & 8                        & 8                                & 8                                                                      & 8                                                                    & 8                             \\
      Cutoff $r_{\text{cut}}$ (\AA) & 6                                                                                                                                                                                                 & 5                & 5                        & 5                                & 5                                                                      & 6                                                                    & 6                             \\
      \hline
      Dataset size                  & \footnote{The LiPS dataset consists of 250001 configurations from an AIMD trajectory.  Subsets of 1000, 1000, and 5000 are randomly sampled for training, validation, and testing, respectively.}
                                    & 1593                                                                                                                                                                                              & 10000            & 10000                    & 10000                            & 10276                                                                  & 387897                                                                                               \\
      Data split                    &                                                                                                                                                                                                   & random           & random                   & random                           & random                                                                 & using~\cite{wen2024an}                                               & using~\cite{kaplan2025matpes} \\
      Split ratio                   &                                                                                                                                                                                                   & 9:0.5:0.5        & 8:1:1                    & 8:1:1                            & 8:1:1                                                                  & 8:1:1                                                                & 8:1:1                         \\
      \hline
      Max epochs                    & 2000                                                                                                                                                                                              & 2000             & 2000                     & 2000                             & 2000                                                                   & 1000                                                                 & 100                           \\
      Learning rate                 & 0.002                                                                                                                                                                                             & 0.001            & 0.002                    & 0.002                            & 0.002                                                                  & 0.001                                                                & 0.0002                        \\
      Batch size                    & 4                                                                                                                                                                                                 & 4                & 20                       & 20                               & 20                                                                     & 20                                                                   & 32                            \\

      EMA decay                     & 0.999                                                                                                                                                                                             & 0.999            & 0.99                     & 0.99                             & 0.99                                                                   & 0.99                                                                 & 0.999                         \\
      Weight decay                  & $1\times10^{-8}$                                                                                                                                                                                  & $1\times10^{-8}$ & $1\times10^{-8}$         & $1\times10^{-8}$                 & $1\times10^{-8}$                                                       & $1\times10^{-1}$                                                     & 0                             \\
      \hline
      Loss function                 & MSE                                                                                                                                                                                               & MSE              & MSE                      & MSE                              & MSE\footnote{Weights for $\bm\mu$, $\bm\alpha$, $\bm\sigma$ set to 1.} & Huber(10)\footnote{Value in the parentheses is the  Huber $\delta$.} & Huber (0.01)                  \\
      Weight $w_\mathcal{E}$        & 1                                                                                                                                                                                                 & 10               & 1                        & -                                & 1                                                                      & -                                                                    & 1                             \\
      Weight $w_\text{F}$           & 1                                                                                                                                                                                                 & 1                & 1                        & -                                & 1                                                                      & -                                                                    & 1                             \\
      Weight $w_\text{S}$           & -                                                                                                                                                                                                 & -                & -                        & -                                & -                                                                      & -                                                                    & 10                            \\
      \hline
   \end{tabular}
\end{table}

\clearpage

\section{Cartesian Natural Tensor Theory}

In the main text, we have provided a concise summary of the key formulas of the Cartesian natural tensor theory that we have developed and employed in this work.
Here, we present a detailed discussion of the theory.
We also provide a computer program to evaluate the formulas presented here, giving both symbolic and numerical results: \url{https://github.com/wengroup/carnet}.

Let's first define some notation.
\begin{itemize}
   \item A bold uppercase letter with a subscript $n$ denotes a rank-$n$ tensor (e.g., $\mathbf{T}_n, \mathbf{U}_n, \mathbf{X}_n$).
         When the rank is not important for the context, we use a bold uppercase letter without a subscript to denote the tensor (e.g.\ $\mathbf{T}, \mathbf{U}, \mathbf{X}$).

   \item A normal uppercase letter with multiple subscripts denotes the components of a tensor in indicial notation (e.g., $T_{i_1 i_2 \dots i_n}, U_{i_1 i_2 \dots i_n}, X_{i_1 i_2 \dots i_n}$).

   \item $\otimes$ denotes the tensor product between two tensors.
         For example, for a rank-$m$ tensor $\mathbf{A}_m$ and a rank-$n$ tensor $\mathbf{B}_n$, their tensor product $\mathbf{C}_{m+n} = \mathbf{A}_m \otimes \mathbf{B}_n$ is a rank-$(m+n)$ tensor with components
         $
            C_{i_1 i_2 \dots i_m j_1 j_2 \dots j_n} = A_{i_1 i_2 \dots i_m} B_{j_1 j_2 \dots j_n}.
         $

   \item $\mathbf{A}^{\otimes k}$ denotes the $k$-fold tensor product of a tensor $\mathbf{A}$ with itself, i.e., $\mathbf{A}^{\otimes k} = \mathbf{A} \otimes \mathbf{A} \otimes \dots \otimes \mathbf{A}$ ($k$ of $\mathbf{A}$).
   \item $\odot^k$ denotes the rank-$k$ contraction between two tensors.
         For example, for a rank-$m$ tensor $\mathbf{A}_m$ and a rank-$n$ tensor $\mathbf{B}_n$ ($m \ge n$), their rank-$k$ contraction $\mathbf{C}_{m+n-2k} = \mathbf{A}_m \odot^k \mathbf{B}_n$ is a rank-$(m+n-2k)$ tensor with components
         $
            C_{i_1 i_2 \dots i_{m-k} j_1 j_2 \dots j_{n-k}} = A_{i_1 i_2 \dots i_{m-k} p_1 p_2 \dots p_k} B_{j_1 j_2 \dots j_{n-k} p_1 p_2 \dots p_k},
         $
         where repeated indices ($p_1, p_2, \dots, p_k$, here) imply summation over them according to the Einstein summation convention.
\end{itemize}

\subsection*{Natural tensors from a unit vector}

For a unit vector $\hat {\mathbf{r}}$, a natural tensor of rank $n$ from it can be created by first constructing a rank-$n$ symmetric polyadic tensor $\mathbf{U} = \hat{\mathbf{r}}^{\otimes n}= \hat{\mathbf{r}} \otimes \dots \otimes \hat{\mathbf{r}}$ ($n$ of $\hat{\mathbf{r}}$ in the tensor product), and then removing the traces to get the rank-$n$ symmetric traceless natural tensor $\mathbf{V}_n$~\cite{jerphagnon1978description}:
\begin{equation} \label{eq:nt:unit:vector}
   V_{i_1 \dots i_n} = C \sum_{t=0}^{\lfloor n/2 \rfloor} (-1)^t \frac{(2n-2t-1)!!}{(2n-1)!!} \left\{ \delta_{i_1i_2} \dots \delta_{i_{2t-1}i_{2t}} U_{i_{2t+1} \dots i_n \underbrace{\scriptstyle ppqq \dots}_{t \text{ pairs}}} \right\}
\end{equation}
where $C = \frac{(2n-1)!!}{n!}$ is a normalization factor (explained below),
$\lfloor {x} \rfloor$ denotes the largest integer not greater than $x$,
the curly braces $\{\}$ denote full symmetrization achieved by summing over all unique permutations,
each term in $\{\}$ contains $t$ Kronecker deltas, and $t$ pairs of indices from $\mathbf{U}$ are contracted (repeated indices $p,q,\dots$ are summed, following the Einstein summation convention).
The rank-0 and rank-1 natural tensors are 1 (scalar) and $\hat{\mathbf{r}}$ (vector), respectively.
\eref{eq:nt:unit:vector} can also be written as~\cite{lehman1989angular}
\begin{equation} \label{eq:nt:unit:vector:2}
   V_{i_1 \dots i_n} =
   C \sum_{t=0}^{\lfloor n/2 \rfloor} (-1)^t \frac{(2n-2t-1)!!}{(2n-1)!!} \{
   \delta_{i_1i_2} \dots \delta_{i_{2t-1}i_{2t}}
   \hat r_{2t+1}
   \hat r_{2t+2}
   \dots
   \hat r_n
   \},
\end{equation}
given that $\hat{\mathbf{r}}$ is a unit vector.

The normalization constant $C = \frac{(2n-1)!!}{n!}$ is chosen such that the rank-$n$ contraction with an arbitrary unit vector $\hat{\mathbf{b}}$ gives the Legendre polynomial of degree $n$:
\begin{equation} \label{eq:legendre:normalization}
   \mathbf{V}_n  \odot^n \hat{\mathbf{b}}^{\otimes n} = P_n(\hat{\mathbf{r}}\cdot\hat{\mathbf{b}}).
\end{equation}
If $\hat{\mathbf{b}}$ is chosen to be $\hat{\mathbf{r}}$, then it normalizes to 1, i.e.,
\begin{equation} \label{eq:unity:normalization}
   \mathbf{V}_n  \odot^n \hat{\mathbf{r}}^{\otimes n} = 1.
\end{equation}
The normalization factor is optional but convenient in practice.
In our computer implementation, we provide an option to include or exclude $C$.
In the main text, we have omitted the normalization factor for simplicity.

Neither \eref{eq:nt:unit:vector} nor \eref{eq:nt:unit:vector:2} is efficient for numerical evaluation due to the summing over different $t$ and the full symmetrization.
\textbf{We propose an efficient method to obtain natural tensors from a unit vector.}
\eref{eq:nt:unit:vector} can be rewriten as
\begin{equation}
   \mathbf{V}_n = \mathbf{H}_{2n} \odot^n \mathbf{U}_n
   \quad\quad
   V_{k_1 \dots k_n} = H_{k_1 \dots k_n i_1 \dots i_n} U_{i_1 \dots i_n},
\end{equation}
where
\begin{equation} \label{eq:nt:unit:vector:H}
   H_{k_1 \dots k_n i_1 \dots i_n} = C \sum_{t=0}^{\lfloor n/2 \rfloor} (-1)^t
   \frac{(2n-2t-1)!!}{(2n-1)!!}
   \{\delta_{ki}^{\otimes n-2t}\delta_{kk}^{\otimes t} \} \delta_{ii}^{\otimes t},
\end{equation}
in which the superscript $^{\otimes t}$ is a shorthand notation for the $t$-fold tensor product of Kronecker deltas (e.g., $ \delta_{ki}^{\otimes 2} = \delta_{k_1i_1}\delta_{k_2i_2}$ and $\delta_{ii}^{\otimes 2} = \delta_{i_1i_2}\delta_{i_3i_4}$),
and the curly braces $\{\,\}$ denote symmetrization achieved by \emph{summing} over all unique permutations of the indices.

This is equivalent to a rearrangement of the summation and symmetrization in \eref{eq:nt:unit:vector}.
However, it can be much more efficient for numerical evaluation,
since one can precompute $\mathbf{H}$ (both symbolically and numerically) for different $n$ and store them, and then $\mathbf{V}_n$ can be efficiently evaluated by performing a single rank-$n$ contraction between $\mathbf{H}_{2n}$ and $\mathbf{U}_n$.

Here, $\mathbf{H}_{2n}$ plays the same role as the spherical harmonic functions to construct spherical tensors from unit vectors.
Note that \eref{eq:nt:unit:vector:H} applies only to completely symmetric tensors such as $\mathbf{U}$; a generalization to arbitrary symmetries is given in \hyperref[sec:decomp:general]{``Decomposition and reconstruction of physical tensors''}.

Let's give some concrete examples of $\mathbf{H}_{2n}$ up to $n=3$.
As discussed above, the rank-0 and rank-1 natural tensors from a unit vector $\hat{\mathbf{r}}$ are 1 and $\hat{\mathbf{r}}$, respectively, so we have
\begin{equation}
   \begin{aligned}
       & H_0 =  r_{i_1}  \quad\text{for}\ n=0               \\
       & H_{k_1 i_1} = \delta_{k_1i_1} \quad\text{for}\ n=1 \\
   \end{aligned}.
\end{equation}
For $n=2$, \eref{eq:nt:unit:vector:H} is
\begin{equation}
   \begin{aligned}
      H_{k_1 k_2 i_1 i_2}
       & = C \sum_{t=0}^{1} (-1)^t \frac{(4-2t-1)!!}{(3)!!} \{\delta_{ki}^{\otimes 2-2t}\delta_{kk}^{\otimes t} \} \delta_{ii}^{\otimes t} \\
       & = C \left(\frac{3!!}{3!!} \{ \delta_{ki}^2\} - \frac{1!!}{3!!} \{ \delta_{kk} \} \delta_{ii} \right)                              \\
       & = C \left(  \{\delta_{k_1 i_1} \delta_{k_2 i_2}\} - \frac{1}{3} \{\delta_{k_1 k_2}\} \delta_{i_1 i_2} \right)                     \\
       & = C \left(  \delta_{k_1 i_1} \delta_{k_2 i_2} - \frac{1}{3} \delta_{k_1 k_2} \delta_{i_1 i_2} \right)            .                \\
   \end{aligned}
\end{equation}
We note that
$\{
   \delta_{k_1 i_1} \delta_{k_2 i_2}\}
   =  \delta_{k_1 i_1} \delta_{k_2 i_2}
$
instead of $\delta_{k_1 i_1} \delta_{k_2 i_2} + \delta_{k_1 i_2} \delta_{k_2 i_1}$, because:
\begin{itemize}
   \item The symmetrization $\{\}$ should be performed only over \textbf{unique} permutations.
   \item $U_{i_1i_2}$ is symmetric, i.e., the two indices $i_1$ and $i_2$ are interchangeable;
\end{itemize}
For $n=3$, \eref{eq:nt:unit:vector:H} evaluates to
\begin{equation}
   H_{k_1 k_2 k_3 i_1 i_2 i_3}
   = C \left(
   \delta_{k_1 i_1} \delta_{k_2 i_2} \delta_{k_3 i_3}
   - \frac{1}{5} \delta_{k_1 i_1} \delta_{k_2 k_3} \delta_{i_2 i_3}
   - \frac{1}{5} \delta_{k_1 i_2} \delta_{k_2 k_3} \delta_{i_1 i_3}
   - \frac{1}{5} \delta_{k_1 i_3} \delta_{k_2 k_3} \delta_{i_1 i_2}
   \right).
\end{equation}

\subsection*{Product of natural tensors}

The tensor product between two natural tensors can be expressed as a direct sum of a set of natural tensors~\cite{coope1970irreducible3,lehman1989angular}.
Given natural tensors $\mathbf{X}_{l_1}$ of rank $l_1$ and $\mathbf{Y}_{l_2}$ of rank $l_2$, their product
\begin{equation} \label{eq:tp}
   \mathbf{Z}_{l_3} = \mathbf{X}_{l_1} \hat\otimes \mathbf{Y}_{l_2}
\end{equation}
is a tensor whose rank lies in the range $|l_1 - l_2| \leq l_3 \leq l_1 + l_2$ (analogous to spherical tensors~\cite{edmonds1996angular}), where $\hat\otimes$ represents the natural tensor product.
For example, when $l_1=1$ and $l_2=2$, $\mathbf{Z}_{l_3}$ is of rank 1, 2, or 3.
The tensor product can be obtained as follows~\cite{lehman1989angular}.

For an even $l_1+l_2-l_3 = 2d$,
\begin{widetext}
   \begin{equation} \label{eq:tp:even}
      \mathbf{Z}_{l_3} = \mathbf{X}_{l_1} \hat\otimes \mathbf{Y}_{l_2}
      =  C_{l_1l_2l_3} \sum_{t=0}^{\min(l_1,l_2)-d} (-1)^t 2^t \frac{(2l_3 - 2t - 1)!!}{(2l_3 - 1)!!} \{\mathbf{X}_{l_1} \odot^{d+t} \mathbf{Y}_{l_2} \otimes \mathbf{I}^{\otimes t}\}
   \end{equation}
   For an odd $l_1+l_2-l_3=2d+1$,
   \begin{equation} \label{eq:tp:odd}
      \mathbf{Z}_{l_3}
      = \mathbf{X}_{l_1} \hat\otimes \mathbf{Y}_{l_2}
      =  D_{l_1l_2l_3} \sum_{t=0}^{\min(l_1,l_2)-d-1} (-1)^t 2^t \frac{(2l_3 - 2t - 1)!!}{(2l_3 - 1)!!} \{ \bm\epsilon : \mathbf{X}_{l_1} \odot^{d+t} \mathbf{Y}_{l_2} \otimes \mathbf{I}^{\otimes t}\},
   \end{equation}
\end{widetext}
where the curly braces $\{\,\}$ denote full symmetrization achieved by summing over all unique permutations,
$\bm\epsilon$ is the Levi-Civita symbol, and
the double colon operator $:$ denotes a double contraction, which is performed between $\bm\epsilon$ and one index of $\mathbf{X}_{l_1}$ and one index of $\mathbf{Y}_{l_2}$.
The normalization factors are given by~\cite{lehman1989angular}
\begin{equation}
   C_{l_1l_2l_3} = \frac{l_1!l_2!(2l_3-1)!!((J_1+1)/2)((J_2+1)/2)!}{l_3!J_1!!J_2!!J_3!!(J/2)!}
\end{equation}
and
\begin{equation}
   D_{l_1l_2l_3} = \frac{2l_1!l_2!(2l_3-1)!!(J_1/2)!(J_2/2)!}{(l_3-1)!(J_1+1)!!(J_2+1)!!(J_3+1)!!((J+1)/2)!},
\end{equation}
where $J = l_1+l_2+l_3$ and $J_i = J - 2l_i - 1$.
The normalization factor $C_{l_1l_2l_3}$ is chosen such that when $\mathbf{X}_{l_1}$ and $\mathbf{Y}_{l_2}$ are constructed from a unit vector $\hat{\mathbf{r}}$ as in \eref{eq:nt:unit:vector}, that is
\begin{equation}
   \mathbf{X}_{l_1} = \mathbf{V}_{l_1} \quad\text{and}\quad
   \mathbf{Y}_{l_2} = \mathbf{V}_{l_2},
\end{equation}
$\mathbf{Z}_{l_3}$ reduces to $\mathbf{V}_{l_3}$ constructed from the same unit vector $\hat{\mathbf{r}}$.
In this case, according to Eqs~\eqref{eq:legendre:normalization} and \eqref{eq:unity:normalization}, we have
\begin{equation}
   \mathbf{Z}_{l_3} \odot^{l_3} \hat{\mathbf{b}}^{\otimes l_3} = P_{l_3}(\hat{\mathbf{r}}\cdot\hat{\mathbf{b}}),
\end{equation}
and
\begin{equation}
   \mathbf{Z}_{l_3} \odot^{l_3} \hat{\mathbf{r}}^{\otimes l_3} = 1.
\end{equation}
The normalization factor $D_{l_1l_2l_3}$ is chosen such that when $\mathbf{X}_{l_1}$ is constructed from a unit vector $\hat{\mathbf{a}}$ and $\mathbf{Y}_{l_2}$ is constructed from a unit vector $\hat{\mathbf{b}}$ as $\hat{\mathbf{b}}$ approaches $\hat{\mathbf{a}}$, we have
\begin{equation}
   \lim_{\hat{\mathbf{b}}\to\hat{\mathbf{a}}}  \frac{ | \mathbf{Z}_{l_3} \odot^{l_3-1} \hat{\mathbf{a}}^{\otimes l_3-1} |} {|\hat{\mathbf{a}}\times\hat{\mathbf{b}}|} = 1,
\end{equation}
where $|\cdot|$ denotes the norm of a vector.

Eqs.~\eref{eq:tp:even} and \eref{eq:tp:odd} are not efficient for numerical evaluation due to the summing over different $t$ and the full symmetrization.
\textbf{We propose an efficient evaluation method as follows.}
They can be rewritten as
\begin{equation}
   \mathbf{Z}_{l_3} = \mathbf{H}_{l_1+l_2+l_3} \odot^{l_1} \mathbf{X}_{l_1} \odot^{l_2} \mathbf{Y}_{l_2}
   \quad
   \quad
   Z_{k_1\dots k_{l_3}} = H_{k_1\dots k_{l_3} i_1\dots i_{l_1} j_1\dots j_{l_2}} X_{i_1\dots i_{l_1}} Y_{j_1\dots j_{l_2}}.
\end{equation}
For even $l_1 + l_2 - l_3=2d$:
\begin{equation} \label{eq:tp:even:H}
   H_{k_1\dots k_{l_3} i_1\dots i_{l_1} j_1\dots j_{l_2}} =
   C_{l_1l_2l_3}
   \sum_{t=0}^{\min(l_1,l_2)-d} (-2)^t
   \frac{(2l_3-2t-1)!!}{(2l_3-1)!!}
   \{ \delta_{ik}^{\otimes l_1 - (d+t)} \delta_{jk}^{\otimes l_2 - (d+t)} \delta_{kk}^{\otimes t} \} \delta_{ij}^{\otimes d+t} .
\end{equation}
For odd $l_1 + l_2 - l_3=2d+1$:
\begin{equation} \label{eq:tp:odd:H}
   H_{k_1\dots k_{l_3} i_1\dots i_{l_1} j_1\dots j_{l_2}} =
   D_{j_1j_2j_3} \sum_{t=0}^{\min(j_1,j_2)-d-1} (-2)^t \frac{(2j_3 - 2t - 1)!!}{(2j_3 - 1)!!}
   \{\epsilon_{kij} \delta_{ik}^{\otimes j_1 - (d+t) - 1 } \delta_{jk}^{\otimes j_2 - (d+t) -1} \delta_{kk}^{\otimes t} \} \delta_{ij}^{\otimes k+t} .
\end{equation}

In computer implementation, one can precompute $\mathbf{H}_{l_1+l_2+l_3}$ (both symbolically and numerically) for different $l_1$, $l_2$, and $l_3$ and store them.
Then $\mathbf{Z}_{l_3}$ can be efficiently evaluated by performing contractions between $\mathbf{H}_{l_1+l_2+l_3}$, $\mathbf{X}_{l_1}$, and $\mathbf{Y}_{l_2}$.

Here, the projection tensor $\mathbf{H}_{l_1+l_2+l_3}$ plays the same role as the Clebsch--Gordan coefficients used in spherical tensor products.

Below we give some concrete examples of $\mathbf{H}$ up to $l_1, l_2, l_3 \leq 2$.

For $l_1 = l_2 = l_3 = 2$, we have $d=1$, and \eref{eq:tp:even:H} evaluates to
\begin{equation}
   \begin{aligned}
      H_{k_1 k_2 i_1 i_2 j_1 j_2}
       & = C_{222}
      \sum_{t=0}^{1} (-2)^t \frac{(3-2t)!!}{3!!} \{ \delta_{ik}^{\otimes 1 - t} \delta_{jk}^{\otimes 1 - t} \delta_{kk}^{\otimes t} \} \delta_{ij}^{\otimes 1+t} \\
       & = C_{222} \left(
      \frac{3!!}{3!!} \{ \delta_{ik}\delta_{jk} \}\delta_{ij}
      - \frac{1!!}{3!!} \{ \delta_{kk} \} \delta_{ij}^{\otimes 2}
      \right)                                                                                                                                                    \\
       & = C_{222} \left(
      \{ \delta_{ik}\delta_{jk} \}\delta_{ij}
      - \frac{1}{3}  \delta_{kk}  \delta_{ij}^{\otimes 2}
      \right)                                                                                                                                                    \\
       & = C_{222}
      \left(
      \delta_{i_1 k_1} \delta_{j_1 k_2} \delta_{i_2 j_2}
      + \delta_{i_1 k_2} \delta_{j_1 k_1} \delta_{i_2 j_2}
      - \frac{1}{3} \delta_{k_1 k_2} \delta_{i_1 j_1} \delta_{i_2 j_2}
      \right)
   \end{aligned}
\end{equation}
We note that $\{ \delta_{ik}\delta_{jk} \}\delta_{ij} = \delta_{i_1 k_1} \delta_{j_1 k_2}\delta_{i_2 j_2} + \delta_{i_1 k_2} \delta_{j_1 k_1} \delta_{i_2 j_2}$
because:
\begin{itemize}
   \item The symmetrization $\{\}$ should be performed only over \textbf{unique} permutations.
   \item $i_1$ and $i_2$ are interchangeable since they are associated with the symmetric natural tensor $\mathbf{X}_2$, and, similarly, $j_1$ and $j_2$ are interchangeable since they are associated with the symmetric natural tensor $\mathbf{Y}_2$.
\end{itemize}
Consequently, we only need to symmetrize over the unique permutations of $k_1$ and $k_2$.

Other examples can be similarly evaluated.
Below we give the results without including the coefficients $C_{l_1l_2l_3}$ and $D_{l_1l_2l_3}$ for simplicity.

\vspace{12pt}
When $l_1 = 0$:

$l_2=0$, $l_3=0$: 1

$l_2=1$, $l_3=1$: $\delta_{k_1 j_1}$

$l_2=2$, $l_3=2$: $\delta_{k_1 j_1}\delta_{k_2 j_2}$

\vspace{12pt}
When $l_1 = 1$:

$l_2=0$, $l_3=1$: $\delta_{k_1 i_1}$

$l_2=1$, $l_3=0$: $\delta_{i_1 j_1}$

$l_2=1$, $l_3=1$: $\epsilon_{k_1 i_1 j_1}$

$l_2=1$, $l_3=2$:
$
   \delta_{k_1i_1}\delta_{k_2 j_1}
   + \delta_{k_1j_1}\delta_{k_2 i_1}
   - \frac{2}{3} \delta_{k_1 k_2} \delta_{i_1 j_1}
$

$l_2=2$, $l_3=1$: $\delta_{k_1 j_1}\delta_{i_1 j_2}$

$l_2=2$, $l_3=2$:
$
   \delta_{k_1 j_2}\epsilon_{k_2 i_1 j_1}
   + \delta_{k_2 j_2}\epsilon_{k_1 i_1 j_1}
$

\vspace{10pt}
When $l_1 = 2$:

$l_2=0$, $l_3=2$: $\delta_{k_1 i_1}\delta_{k_2 i_2}$

$l_2=1$, $l_3=1$: $\delta_{k_1 i_1}\delta_{j_1 i_2}$

$l_2=1$, $l_3=2$: $
   \delta_{k_1 i_2}\epsilon_{k_2 j_1 i_1}
   + \delta_{k_2 i_2}\epsilon_{k_1 j_1 i_1}
$

$l_2=2$, $l_3=0$: $\delta_{i_1 j_1}\delta_{i_2 j_2}$

$l_2=2$, $l_3=1$: $\epsilon_{k_1i_1 j_1} \delta_{i_2 j_2}$

$l_2=2$, $l_3=2$:
$
   \delta_{i_1 k_1} \delta_{j_1 k_2} \delta_{i_2 j_2}
   + \delta_{i_1 k_2} \delta_{j_1 k_1} \delta_{i_2 j_2}
   - \frac{1}{3} \delta_{k_1 k_2} \delta_{i_1 j_1} \delta_{i_2 j_2}
$

\phantomsection
\subsection*{Decomposition and reconstruction of physical tensors}
\label{sec:decomp:general}

An approach to obtaining the decomposition spectrum of high-rank tensors was proposed by Coope and coworkers~\cite{coope1965irreducible,coope1970irreducible2}.
\textbf{We extend their approach and propose a systematic procedure for obtaining the decomposition spectrum of physical tensors of arbitrary rank and symmetry.
}
We extend their work in two aspects:
first, we propose to use QR factorization to find linearly independent and orthogonal natural tensor candidates;
and second, we introduce a symmetry-informed elimination procedure to identify natural tensor candidates that are compatible with the symmetry of the physical tensor.

\subsubsection*{Decomposition spectrum of Cartesian tensors}

We first briefly review the method by Coope and coworkers~\cite{coope1965irreducible,coope1970irreducible2} to obtain the decomposition spectrum of generic Cartesian tensors.
A rank-$n$ Cartesian tensor $\mathbf{T}_n$ can be decomposed into the direct sum of irreducible components (natural tensors) as
\begin{equation}
   \mathbf{T}_n = \sum_{\oplus m, p}  \mathbf{X}_m^p,
\end{equation}
where $\mathbf{X}_m^p$ is a natural tensor of rank $m$ (called the \emph{weight}), and $p$ (called the \emph{seniority}) is used to distinguish different natural tensors of the same rank~\cite{jerphagnon1978description}.
This means there can be multiple natural tensors of the same rank $j$ in the decomposition spectrum of a rank-$n$ Cartesian tensor.
\tref{tab:reduction:spectrum} lists the decomposition spectrum of generic Cartesian tensors up to rank seven.

\begin{table}[h!]
   \caption[Decomposition spectrum of Cartesian tensors]{Decomposition spectrum of generic Cartesian tensors up to rank seven.
      Number of independent irreducible components in the decomposition spectrum of a rank-$n$ tensor for different weight $m$.
      For example, a rank-2 Cartesian tensor can be decomposed into a rank-0, rank-1, and a rank-2 natural tensors; a rank-3 Cartesian tensor can be decomposed into a rank-0, three rank-1, two rank-2, and a rank-3 natural tensors.
      The table is based on Figure~1 in \olcite{coope1965irreducible}.
   }
   \label{tab:reduction:spectrum}
   \newcolumntype{C}{w{c}{18pt}}  
   \begin{tabular}{|CC|CCCCCCCC|}
      \hline
                                                  &   & \multicolumn{7}{c}{Rank $n$} &                               \\
                                                  &   & 0                            & 1 & 2 & 3 & 4 & 5  & 6  & 7   \\
      \hline
      \multirow{8}{*}{\rotatebox{90}{Weight $m$}} & 0 & 1                            &   & 1 & 1 & 3 & 6  & 15 & 36  \\
                                                  & 1 &                              & 1 & 1 & 3 & 6 & 15 & 36 & 91  \\
                                                  & 2 &                              &   & 1 & 2 & 6 & 15 & 40 & 105 \\
                                                  & 3 &                              &   &   & 1 & 3 & 10 & 29 & 84  \\
                                                  & 4 &                              &   &   &   & 1 & 4  & 15 & 49  \\
                                                  & 5 &                              &   &   &   &   & 1  & 5  & 21  \\
                                                  & 6 &                              &   &   &   &   &    & 1  & 6   \\
                                                  & 7 &                              &   &   &   &   &    &    & 1   \\
      \hline
   \end{tabular}
\end{table}

To obtain the natural tensors, one first constructs the \emph{natural projector},
\begin{equation} \label{eq:natural:projector}
   \mathbf{E}_{2m} = \sum_{t=0}^{\lfloor m /2 \rfloor} c_t \mathbf{I}_{ik}^{\otimes(m-2t)} \mathbf{I}_{ii}^{\otimes t} \mathbf{I}_{kk}^{\otimes t},
\end{equation}
that is,
\begin{equation} \label{eq:Eki}
   E_{k_1\dots k_m i_1\dots i_m}  = \sum_{t=0}^{\lfloor m /2 \rfloor} c_t (\delta_{ik})^{m-2t} (\delta_{ii})^t( \delta_{kk})^t
\end{equation}
$\mathbf{E}_{2m}$ is a rank-$2m$ tensor that maps between the space of natural tensors and the space of ordinary Cartesian tensors.
Following the notation in \olcite{coope1965irreducible}, $(\delta_{rr})^t$ is a shorthand notation for a full symmetrization of the product of $t$ Kronecker deltas, each contracting a pair of indices.
For example,
\begin{equation}
   (\delta_{ik})^2 = \frac{1}{2} (
   \delta_{i_1k_1}\delta_{i_2k_2}
   + \delta_{i_1k_2}\delta_{i_2k_1}
   ),
\end{equation}
\begin{equation}
   (\delta_{ii})^2 = \frac{1}{3} (
   \delta_{i_1i_2}\delta_{i_3i_4}
   + \delta_{i_1i_3}\delta_{i_2i_4}
   + \delta_{i_1i_4}\delta_{i_2i_3}
   ),
\end{equation}
and when $m=3$ and $t=1$, we have:
\begin{equation}
   \begin{aligned}
      \delta_{ik} \delta_{ii} \delta_{kk}
      = & \frac{1}{9} (
      \delta_{i_1k_1} \delta_{i_2i_3} \delta_{k_2k_3}
      + \delta_{i_1k_2} \delta_{i_2i_3} \delta_{k_1k_3}
      + \delta_{i_1k_3} \delta_{i_2i_3} \delta_{k_1k_2}     \\
        & + \delta_{i_2k_1} \delta_{i_1i_3} \delta_{k_2k_3}
      + \delta_{i_2k_2} \delta_{i_1i_3} \delta_{k_1k_3}
      + \delta_{i_2k_3} \delta_{i_1i_3} \delta_{k_1k_2}     \\
        & + \delta_{i_3k_1} \delta_{i_1i_2} \delta_{k_2k_3}
      + \delta_{i_3k_2} \delta_{i_1i_2} \delta_{k_1k_3}
      + \delta_{i_3k_3} \delta_{i_1i_2} \delta_{k_1k_2}
      ).
   \end{aligned}
\end{equation}
The coefficient $c_t$ is defined as:
\begin{equation} \label{eq:ct}
   c_t = (-1)^t \frac{(m!)^2}{(2m)!} \binom{m}{t} \binom{2m-2t}{m},
\end{equation}
where $\binom{a}{b}$ is the binomial coefficient.

\eref{eq:natural:projector} can essentially be used to project out the rank $m$ natural tensor from a rank-$m$ Cartesian tensor.
When $t=0$, it symmetrizes the Cartesian tensor, and when $t>0$, it removes the traces.

The projector $\mathbf{G}^p_{m+n}$ can be seen as a generalization of $\mathbf{H}_{2n}$ in \eref{eq:nt:unit:vector:H} to tensors with arbitrary symmetries: \eref{eq:nt:unit:vector:H} applies only to completely symmetric tensors such as $\mathbf{U}$, whereas $\mathbf{G}^p_{m+n}$ additionally handles the symmetrization of general tensors (via the $t=0$ contribution in $\mathbf{E}_{2m}$) before removing traces.
Since $\mathbf{U}$ is already fully symmetric, only trace removal is needed, making \eref{eq:nt:unit:vector:H} more concise.

With $\mathbf{E}_{2m}$, we can also construct the projectors that map between a general tensor $\mathbf{T}_n$ and the natural tensor $\mathbf{X}_m^p$ of weight $m$.
First, the projector tensor $\mathbf{G}^p_{m+n}$ that embeds a natural tensor $\mathbf{X}_m^p$ into a Cartesian tensor space of rank $n$ is constructed as follows.

For even $n-m$,
\begin{equation} \label{eq:G:even}
   \mathbf{G}^p_{m+n} = \mathbf{E}_{2m} \otimes \mathbf{I}^{\otimes (n-m)/2}
   \quad\quad
   G_{k_1 \dots k_m i_1 \dots i_n}^p = E_{k_1 \dots k_m i_1 \dots i_m}
   (\delta_{ii})^{(n-m)/2},
\end{equation}

For odd $n-m$,
\begin{equation} \label{eq:Gnj:odd}
   \mathbf{G}^p_{m+n} = \mathbf{E}_{2m} \odot \bm \epsilon \otimes \mathbf{I}^{\otimes (n-m-1)/2}
   \quad
   G_{k_1 \dots k_m i_1 \dots i_n}^p = E_{k_1 \dots k_m i_1 \dots i_m}
   \epsilon_{i_{m} i_{m+1} i_{m+2}} (\delta_{ii})^{(n-m-1)/2}.
\end{equation}

$\mathbf{G}_{m+n}^p$ has the following property:
\begin{equation} \label{eq:G:orthogonal}
   \mathbf{G}^p_{m+n} \odot^n \mathbf{G}^{q}_{m+n} = g_{pq} \mathbf{E}_{2m},
\end{equation}
where $g_{pq}$ is a scalar.
All the values of $g_{pq}$ of different seniority $p$ and $q$ form a symmetric matrix $\mathbf{g}$, and its inverse is
\begin{equation} \label{eq:h}
   \mathbf{h}  = \mathbf{g}^{-1}.
\end{equation}

With these, we can construct the natural tensor extractor $\mathbf{H}^p_{(m|n)}$, which extracts the natural tensor $\mathbf{X}_m^p$ from a Cartesian tensor $\mathbf{T}_n$.
It is defined as
\begin{equation} \label{eq:H:extractor}
   \mathbf{H}^p_{m+n} = \sum_{q} h_{pq} \mathbf{G}^q_{m+n}.
\end{equation}

The above discussion concludes the construction of the projector tensors $\mathbf{G}^p_{m+n}$ and $\mathbf{H}^p_{m+n}$ to decompose and reconstruct natural tensors of weight $m$ from a rank-$n$ Cartesian tensor.
Let us reiterate the key results: \\
\textbf{
   Given a rank-$n$ Cartesian tensor $\mathbf{T}_n$, we can extract the natural tensor $\mathbf{X}_m^p$ of rank $m$ as
   \begin{equation} \label{eq:extract}
      \mathbf{X}^p_m = \mathbf{H}^p_{m+n} \odot^n \mathbf{T}_n;
   \end{equation}
   Given a natural tensor $\mathbf{X}_m$ of weight $m$, we can embed it into a tensor space of rank-$n$ by
   \begin{equation} \label{eq:embed}
      \mathbf{S}^p_n = \mathbf{G}^p_{m+n} \odot^m \mathbf{X}_m;
   \end{equation}
   Consequently, a rank-$n$ Cartesian tensor $\mathbf{T}_n$ can be reconstructed from its set of natural representations ${\mathbf{X}_m^p}$ as
   \begin{equation} \label{eq:reconstruct}
      \mathbf{T}_n = \sum_{m, p} \mathbf{G}^p_{m+n} \odot^m \mathbf{X}_m^p.
   \end{equation}
}

\phantomsection
\subsubsection*{Selection of independent natural tensors}
\label{sec:ind:natural:tensors}

The above procedure provides a way to construct the projector tensors $\mathbf{G}^p_{m+n}$ and $\mathbf{H}^p_{m+n}$, which can then be used to decompose a Cartesian tensor into natural tensors (\eref{eq:extract}), and, conversely, reconstruct a Cartesian tensor from its natural representations (\eref{eq:reconstruct}).
The number of candidate projector tensors constructed this way, however, is often larger than the actual number of independent natural tensors.
For example, for a rank-3 Cartesian tensor, the above procedure gives three candidates of rank-2 projectors, but only two of them are independent as seen in \tref{tab:reduction:spectrum}.
Explicitly, when $n=3$ and $m=2$, from \eref{eq:Gnj:odd}, we have $\mathbf{G}_5 = \mathbf{E}_4 \odot \bm\epsilon$.
There are three candidate projectors:
\begin{equation}
   \begin{aligned}
       & G_{k_1k_2i_1i_2i_3}^1 = E_{k_1 k_2 i_1 i_\alpha} \epsilon_{i_\alpha i_2 i_3} \\
       & G_{k_1k_2i_1i_2i_3}^2 = E_{k_1 k_2 i_1 i_\alpha} \epsilon_{i_2 i_\alpha i_3} \\
       & G_{k_1k_2i_1i_2i_3}^3 = E_{k_1 k_2 i_1 i_\alpha} \epsilon_{i_2 i_3 i_\alpha} \\
   \end{aligned}
   ,
\end{equation}
depending on which index $i_\alpha$ of the Levi-Civita symbol $\bm\epsilon$ is contracted with the natural projector $\mathbf{E}_4$.
But they are linearly dependent, and only two of them are independent.
We can select any two of them as the independent projectors, and the third one can be expressed as a linear combination of the two independent ones.

The identification of linearly independent $\mathbf{G}^p_{m+n}$ ($p$ is an index for different candidates) can be done as follows:
\begin{enumerate}
   \item Create an arbitrary natural tensor $\mathbf{X}_m$ of weight $m$;
   \item Embed $\mathbf{X}_m$ into the space of rank-$n$ Cartesian tensors via \eref{eq:embed} for all candidates.
   \item Find a set of linearly independent $\mathbf{S}^p_n$, and the corresponding $\mathbf{G}^p_{m+n}$ are the independent projectors.
\end{enumerate}
Step 3 above can be conducted using the QR factorization~\cite{golub2013matrix}.
Specifically, we represent the tensor $\mathbf{S}^p_n$ as a vector $\mathbf{s}^p$ of length $3^n$ and construct a $3^n$ by $p$ matrix $\mathbf{A}$ by putting each $\mathbf{s}^p$ as a column.
Then, a QR~factorization of $\mathbf{A}$ is performed to identify its linearly independent columns, and the corresponding $\mathbf{G}^p_{m+n}$ are the independent projectors.

\subsubsection*{Decomposition of symmetric Cartesian tensors}

\begin{table}[h!]
   \caption[Decomposition spectrum of physical tensors]{Decomposition spectrum of the physical tensors with inherent symmetry.
      Indices in a pair of parentheses are symmetric, and those in a pair of double parentheses are symmetric with respect to exchange of the two pairs.
   }
   \label{tab:reduction:physical:tensors}
   \begin{tabular}{cccccccccc}
      \hline
      Rank $n$ & Symmetry   & Example                & \# independent components & \multicolumn{6}{c}{Spectrum}                     \\
               &            &                        &                           & $l$:                         & 0 & 1 & 2 & 3 & 4 \\
      \hline
      0        &            & Energy                 & 1                         &                              & 1                 \\
      1        & i          & Dipole moment / Forces & 3                         &                              &   & 1             \\
      2        & ij         & Nuclear shielding      & 9                         &                              & 1 & 1 & 1         \\
               & (ij)       & Polarizability         & 6                         &                              & 1 &   & 1         \\
      4        & ijkl       & Generic rank-4 tensor  & 81                        &                              & 3 & 6 & 6 & 3 & 1 \\
               & ((ij)(kl)) & Elasticity             & 21                        &                              & 2 &   & 2 &   & 1 \\
      \hline
   \end{tabular}
\end{table}

So far, we have discussed the decomposition of generic Cartesian tensors.
Many physical tensors, however, have inherent symmetries.
For example, the polarizability is a symmetric rank-2 tensor, and its decomposition spectrum contains only a rank-0 and a rank-2 natural tensor (\tref{tab:reduction:physical:tensors}), without the rank-1 part for a generic rank-2 tensor (\tref{tab:reduction:spectrum}).

\textbf{For tensors with inherent symmetries, the number of independent natural tensors in their decomposition spectrum is fewer than that of generic Cartesian tensors.
Thus, the number of projectors $\mathbf{G}^p_{m+n}$ and $\mathbf{H}^p_{(m|n)}$ should be reduced accordingly.}
Here, we propose a systematic procedure to identify the independent projectors for physical tensors with inherent symmetries.

\begin{enumerate}

   \item \textbf{ Obtain all independent $\mathbf{G}^p_{m+n}$ as discussed in \hyperref[sec:ind:natural:tensors]{``Selection of independent natural tensors''}, where $p$ is an index for different candidates ($p=1,2,\dots,N$). }

   \item \textbf{Identify a subset of $N_g$ unique linearly independent mapping tensors from all $N$ candidates $\mathbf{G}^1, \ldots, \mathbf{G}^N$.}
         Create a rank-$n$ tensor $\mathbf{T}_n$ with the target symmetry, and then compute $\mathbf{F}_n^p = \mathbf{G}^p_{m+n}\odot^n \mathbf{T}_n$ for all independent $\mathbf{G}^p_{m+n}$.
         Because there is inherent symmetry in $\mathbf{T}_n$, some $\mathbf{F}_n^p$ will be identical to others.
         Next, group the $\mathbf{G}^p_{m+n}$ according to $\mathbf{F}_n^p$: the $\mathbf{G}^p_{m+n}$ that give the same $\mathbf{F}_n^p$ are equivalent and are put in the same group.
         Let $N_g$ be the number of groups; then $N_d = N - N_g$ is the number of dependent $\mathbf{G}^p_{m+n}$.

   \item \textbf{Construct the decomposition projector $\mathbf{H}$ to extract natural tensors from a physical tensor with inherent symmetry. }
         Compute $\mathbf{H}^p_{m+n} = \sum_q h_{pq} \mathbf{G}^q_{m+n}$ (i.e., \eref{eq:H:extractor}) for $p=1,2,\dots,N$, where $q=1,2,\dots,N$.
         In total, there are $N$ of $\mathbf{H}^p_{m+n}$, but only $N_g$
         of them are independent, since they are linear combinations of $\mathbf{G}^p_{m+n}$ and there are only $N_g$ unique $\mathbf{G}^p_{m+n}$ as identified in step 2.
         Without loss of generality, let the first $N_g$ of $\mathbf{H}^p_{(j|n)}$ be the independent ones, and the remaining $N_d$ of them be the dependent ones (If not, we can reindex them).
         Then, we can express the dependent $\mathbf{H}^q_{m+n}$ as linear combinations of the independent ones:
         $\mathbf{H}^q_{m+n} = \sum_{p=1}^{N_g} \beta_{qp} \mathbf{H}^p_{m+n}$, where $q=N_g+1,N_g+2,\dots,N$.
         The coefficients $\beta_{qp}$ can be solved by combining this equation and \eref{eq:H:extractor}.

   \item \textbf{Construct the reconstruction tensor $\mathbf{Q}$ to embed natural tensors into the space of physical tensor with inherent symmetry. }
         We create
         $\mathbf{Q}^p_{m+n} = \mathbf{G}^p_{m+n} + \sum_{q=N_g+1}^{N} \beta_{qp} \mathbf{G}^q_{m+n}$, where $p=1,2,\dots,N_g$.
\end{enumerate}

With $\mathbf{H}^p_{m+n}$ obtained in step 3 and $\mathbf{Q}^p_{m+n}$ obtained in step 4  (where $p=1,2,\dots,N_g$), one can use Eqs.~\eqref{eq:extract}, \eqref{eq:embed}, and \eqref{eq:reconstruct} to extract the natural tensors from a physical tensor with inherent symmetry, and embed the natural tensors into the space of physical tensor with inherent symmetry.
Consequently, this allows for the mapping between natural tensors and physical tensors of arbitrary rank and with any inherent symmetry.

\subsubsection*{Example for polarizability}

Using the polarizability tensor as an example, we illustrate the above procedure to identify the independent projectors for a symmetric rank-2 tensor.
As shown in \tref{tab:reduction:spectrum}, a generic rank-2 Cartesian tensor can be decomposed into a rank-0, a rank-1, and a rank-2 natural tensors, but a symmetric rank-2 tensor does not have the rank-1 part in its decomposition spectrum, and its decomposition spectrum only consists of a rank-0 and a rank-2 natural tensors (\tref{tab:reduction:physical:tensors}).

Let's get the projector to extract the rank-0 natural tensor.

Step 1, using \eref{eq:G:even}, when $n=2$ and $m=0$, we have
\begin{equation}
   G^1_{i_1i_2} = E_0 \delta_{i_1 i_2} = 1\cdot\delta_{i_1 i_2}=\delta_{i_1 i_2}.
\end{equation}

Step 2, given that there is a single candidate $G$, it is independent by default, and thus $N_g=1$ and $N_d=0$.

Step 3, $\mathbf{G}^1 \odot^2 \mathbf{G}^1 = \delta_{i_1 i_2} \delta_{i_1 i_2} = 3 $.
Then according to \eref{eq:G:orthogonal} and \eref{eq:h}, we have $g_{11} = 3$ and $h_{11} = g_{11}^{-1} = 1/3$.
Therefore, according to \eref{eq:H:extractor}, we have
$H^1_{i_1i_2} = h_{11} G^1_{i_1i_2} = \frac{1}{3} \delta_{i_1 i_2}$.

Step 4, given that there is a single candidate $G$,
we have $Q^1_{i_1 i_2} = G^1_{i_1 i_2} = \delta_{i_1 i_2}$.

\vspace{20pt}
Next, let's get the projector to extract the rank-2 natural tensor.

Step 1, using \eref{eq:G:even}, when $n=2$ and $m=2$, we have
\begin{equation}
   \begin{aligned}
      G_{k_1 k_2 i_1 i_2}
       & = E_{k_1 k_2 i_1 i_2}
      =
      \sum_{t=0}^1 c_t (\delta_{ik})^{2-2t} (\delta_{ii})^t( \delta_{kk})^t \\
       & = c_0 (\delta_{ik})^2 + c_1 (\delta_{ii}) (\delta_{kk})
      = \frac{1}{2} ( \delta_{i_1k_1} \delta_{i_2k_2} + \delta_{i_1k_2} \delta_{i_2k_1} )
      - \frac{1}{3} \delta_{i_1 i_2} \delta_{k_1 k_2}                       \\\end{aligned} \end{equation} where in the second equation, we have used \eref{eq:Eki}, and $c_0=1$ and $c_1=-\frac{1}{3}$ are computed using the formula of $c_t$ in \eref{eq:ct}.

Step 2, given that there is a single candidate $G$, it is independent by default, and thus $N_g=1$ and $N_d=0$.

Step 3, $\mathbf{G}^1 \odot^2 \mathbf{G}^1 = G_{k_1 k_2 i_1 i_2} G_{k_3 k_4 i_1 i_2} =
      [ \frac{1}{2} ( \delta_{i_1k_1} \delta_{i_2k_2} + \delta_{i_1k_2} \delta_{i_2k_1} ) - \frac{1}{3} \delta_{i_1 i_2} \delta_{k_1 k_2} ]
   \cdot
   [ \frac{1}{2} ( \delta_{i_1k_3} \delta_{i_2k_4} + \delta_{i_1k_4} \delta_{i_2k_3} ) - \frac{1}{3} \delta_{i_1 i_2} \delta_{k_3 k_4} ]
   =
   [ \frac{1}{2} ( \delta_{k_1k_3} \delta_{k_2k_4} + \delta_{k_1k_4} \delta_{k_2k_3} ) - \frac{1}{3} \delta_{k_1 k_2} \delta_{k_3 k_4} ]
$.
Then according to \eref{eq:G:orthogonal} and \eref{eq:h}, we have $g_{11} = 1$ and $h_{11} = g_{11}^{-1} = 1$.
Then according to \eref{eq:H:extractor}, we have $H^1_{k_1 k_2 i_1 i_2} = h_{11} G^1_{k_1 k_2 i_1 i_2} = G^1_{k_1 k_2 i_1 i_2}$ .

Step 4, given that there is a single candidate $G$, we have $Q^1_{k_1 k_2 i_1 i_2} = G^1_{k_1 k_2 i_1 i_2}$.

\clearpage
\section{Comparison with other Cartesian methods}

There are several existing equivariant Cartesian methods for atomistic ML.
According to the way the internal equivariant features are constructed, they can be categorized into two groups: those that use reducible Cartesian tensors as features (including MTP~\cite{shapeev2016moment}, FieldSchNet~\cite{schutt2021equivariant}, CAMP~\cite{wen2025cartesian}, and HotPP~\cite{wang2024equivariant}) and those that use irreducible Cartesian tensors as features (including REANN~\cite{zhang2021physically}, CACE~\cite{cheng2024cartesian}, TensorNet~\cite{simeon2023tensornet}, ICTP~\cite{zaverkin2024higher}, and TACE~\cite{xu2025tace}).

Methods using reducible Cartesian tensors, such as CAMP and HotPP, typically operate on full Cartesian products that satisfy $E(3)$ equivariance but cannot respect the intrinsic symmetries inherent in many physical tensors.
As a concrete example, a rank-4 elastic constant tensor $C_{ijkl}$ contains $3^4 = 81$ components, yet it possesses only 21 independent components once minor and major symmetries ($C_{ijkl} = C_{jikl} = C_{klij}$) are considered.
While models like HotPP must parameterize and model all 81 components, CarNet automatically utilizes irreducible Cartesian tensors to work directly within the symmetry-allowed manifold.
By focusing only on the 21 independent components, CarNet eliminates mathematical redundancy and achieves a more compact representation without sacrificing geometric expressiveness.

Existing methods using irreducible Cartesian tensors address symmetry issues but lack the generality to systematically handle tensors of arbitrary ranks and symmetries.
REANN and CACE focus exclusively on scalar targets.
TensorNet advances this by modeling targets up to rank-2, yet its internal features are restricted to rank-0, 1, and 2, with no systematic method provided for higher-rank extension.
While ICTP and the recent TACE are more similar to CarNet in their use of irreducible Cartesian internal features, they face distinct limitations.
Compared to CarNet, ICTP relies on inefficient loops to construct natural tensors and perform tensor products (i.e., Eqs.~\eqref{eq:nt:unit:vector:2}, \eqref{eq:tp:even}, and \eqref{eq:tp:odd}), whereas CarNet utilizes an efficient projector-based approach.
Moreover, ICTP and TACE lack a systematic framework to identify independent natural tensors or construct projectors for high-rank targets with complex inherent symmetries.
In principle, extending these methods to arbitrary ranks is possible, but no formal theory or practical implementation for handling high-rank tensors with complex inherent symmetries currently exists.
Consequently, these models have only been demonstrated on targets up to the straightforward rank-2, where symmetry constraints are relatively straightforward to manage.
In contrast, this work establishes a general Cartesian model based on three essential components: (1) constructing natural tensors from unit vectors, (2) performing tensor products between natural tensors, and (3) decomposing and reconstructing physical tensors of arbitrary rank with any inherent symmetry.
This third component fills the gap existing in ICTP and TACE.

\clearpage
\section{Nuclear chemical shifts}
\label{sec:chemical:shift}

The nuclear chemical shifts can be obtained from the nuclear shielding tensor as
\begin{equation}
   \delta = \sigma_\text{ref}^Z - \frac{1}{3} \text{Tr}[\bm\sigma],
\end{equation}
where Tr denotes the trace of a rank-2 tensor, and $\sigma_\text{ref}^Z$ is the
reference shifts for a tetramethylsilane molecule.
The reference shifts computed with the PBE0/def2-TZVP functional are $\sigma^\text{H}$ = 31.77~ppm and $\sigma^\text{C}$ = 188.53~ppm~\cite{gastegger2021machine}.

The nuclear shielding tensor $\bm\sigma$ in the original ethanol dataset~\cite{gastegger2021machine} is originally given in atomic units (a.u.).
To convert it to ppm, a conversion factor $c$ is multiplied:
\begin{equation}
   c = \alpha^2 \times 10^6 =53.25135,
\end{equation}
where
\begin{equation}
   \alpha = \frac{e^2}{4\pi \epsilon_0 \hbar c} \approx 0.007297
\end{equation}
is the fine-structure constant.

\setlength{\tabcolsep}{4pt}
\begin{table*}[tbh!]
   \centering
   \caption[Nuclear shielding of ethanol]{Additional results on learning nuclear shielding of ethanol.
      Mean absolute errors (MAEs) of the nuclear shielding tensor for all elements $\bm \sigma_{\text{all}}$, hydrogen $\bm \sigma_\text{H}$, carbon $\bm \sigma_\text{C}$, and oxygen $\bm \sigma_\text{O}$,
      as well as the corresponding chemical shift  $\delta_{\text{all}}$, $\delta_\text{H}$, $\delta_\text{C}$, and $\delta_\text{O}$ are reported.
      Chemical shift is computed as one third of the trace of the shielding tensor, with a constant offset.
   }
   \label{tab:ethanol:vacuum}
   \begin{tabular}{lccccccccc}
      \hline
                  & $\bm \sigma_\text{all}$
                  & $\bm \sigma_\text{H}$
                  & $\bm \sigma_\text{C}$
                  & $\bm \sigma_\text{O}$
                  & $\delta_{\text{all}}$   & $\delta_\text{H}$ & $\delta_\text{C}$ & $\delta_\text{O}$                                 \\
      \hline
      FieldSchNet &                         &                   &                   &                   & 0.169 & 0.123 & 0.194 & 0.401 \\
      TensorNet   &                         &                   &                   &                   & 0.139                         \\
      \model      & 0.047                   & 0.021             & 0.066             & 0.166             & 0.031 & 0.013 & 0.048 & 0.105 \\
      \hline
   \end{tabular}
\end{table*}

\clearpage
\section{Tensor product modes}

For a tensor product $\mathbf{Z}_{l_3} = \mathbf{X}_{l_1}\hat\otimes \mathbf{Y}_{l_2}$, different combinations of $l_1$ and $l_2$ can lead to the same output tensor of rank $l_3$.
The 3-tuple $(l_1, l_2, l_3)$ is called a path.
According to the product rule of natural tensors~\, for a given $l_1$ and $l_2$, all $l_3$ satisfying
\begin{equation} \label{eq:tp:rule}
   |l_1-l_2 | \leq l_3 \leq l_1+l_2
\end{equation}
are allowed values.
In building the models, we restrict the maximum rank of the used tensors to $L$, meaning that
\begin{equation} \label{eq:tp:max:L}
   0 \leq l_1, l_2, l_3 \leq L.
\end{equation}

\textbf{`Full' mode}.
All paths $(l_1, l_2, l_3)$ satisfying Eqs.~\eqref{eq:tp:rule} and \eqref{eq:tp:max:L} are allowed.

\textbf{`Lite' mode}.
In addition to Eqs.~\eqref{eq:tp:rule} and \eqref{eq:tp:max:L}, the paths are further restricted to those with
\begin{equation} \label{eq:tp:lite}
   l_3 = |l_1 - l_2|.
\end{equation}
This is inspired by the tensor contraction rule in the CAMP model~\cite{wen2025cartesian}, where all indices in $\mathbf{X}_{l_1}$ are contracted away by the indices in $\mathbf{Y}_{l_2}$.
It is required that $l_1 \leq l_2$, and the output is
$
   \mathbf{Z}_{l_2-l_1} = \mathbf{X}_{l_1} \odot^{l_1} \mathbf{Y}_{l_2}
$, where $\odot^{l_1}$ denotes an $l_1$-th order contraction.
As can be seen, this operation is not symmetric, since the rank $l_1$ has to be smaller than the rank $l_2$; in other words, the order of $\mathbf{X}_{l_1}$ and $\mathbf{Y}_{l_2}$ matters.
Here, we improve this by allowing for a symmetric operation according to \eref{eq:tp:lite}, which allows for both $l_1 \leq l_2$ and $l_1 \geq l_2$.

\textbf{`Level' mode}.
Eqs.~\eqref{eq:tp:rule} and \eqref{eq:tp:max:L} are used for scalar output $l_3=0$.
But for $l_3>0$, in addition to the two equations, the paths are further restricted to those with
\begin{equation}
   l_1 + l_2 \leq l_l,
\end{equation}
where $l_l$ is an integer in the range of $[L, 2L]$.
This mode provides a way to control the complexity of the model by tuning the level $l_l$.
When $l_l$ is chosen to have a value of $2L$, this is equivalent to the `full' mode.
For other values, it further restricts the tensor products to be between tensors of smaller ranks.
\textbf{We use $l_l=L$ for all results reported in this work.}

Supplementary Tables~\ref{tab:tp:path:L2}, \ref{tab:tp:path:L3}, and \ref{tab:tp:path:L4} explicitly list the paths for the `full', `lite', and `level' models for $L=2$, $L=3$, and $L=4$, respectively.
It should be clear that:
\begin{enumerate}
   \item The full mode has many more paths for larger $l_3$ than for smaller $l_3$, particularly when $L$ is large.
   \item The lite mode has roughly the same number of paths for different $l_3$.
         This is also the case for the level mode.
   \item The total number of paths for the lite mode and the level model are about the same.
\end{enumerate}

\begin{table}[h]
   \centering
   \caption{Tensor product paths for $L=2$}
   \label{tab:tp:path:L2}
   \setlength{\tabcolsep}{6pt} 
   \begin{tabular}{cccl}
      \hline
      \textbf{Mode} & \textbf{$l_3$} & \textbf{\# Paths} & \textbf{Paths} $(l_1, l_2, l_3)$                                 \\
      \hline
      \multirow{3}{*}{full}
                    & 0              & 3                 & (0, 0, 0), (1, 1, 0), (2, 2, 0)                                  \\
                    & 1              & 6                 & (0, 1, 1), (1, 0, 1), (1, 1, 1), (1, 2, 1), (2, 1, 1), (2, 2, 1) \\
                    & 2              & 6                 & (0, 2, 2), (1, 1, 2), (1, 2, 2), (2, 0, 2), (2, 1, 2), (2, 2, 2) \\
      \hline
      \multirow{3}{*}{lite}
                    & 0              & 3                 & (0, 0, 0), (1, 1, 0), (2, 2, 0)                                  \\
                    & 1              & 4                 & (0, 1, 1), (1, 0, 1), (1, 2, 1), (2, 1, 1)                       \\
                    & 2              & 2                 & (0, 2, 2), (2, 0, 2)                                             \\
      \hline
      \multirow{3}{*}{level}
                    & 0              & 3                 & (0, 0, 0), (1, 1, 0), (2, 2, 0)                                  \\
                    & 1              & 3                 & (0, 1, 1), (1, 0, 1), (1, 1, 1)                                  \\
                    & 2              & 3                 & (0, 2, 2), (1, 1, 2), (2, 0, 2)                                  \\
      \hline
   \end{tabular}
\end{table}

\begin{table}[h]
   \centering
   \caption{Tensor product paths for $L=3$}
   \label{tab:tp:path:L3}
   \setlength{\tabcolsep}{6pt}
   \small
   \begin{tabular}{cccl}
      \hline
      \textbf{Mode} & \textbf{$l_3$} & \textbf{\# Paths} & \textbf{Paths} $(l_1, l_2, l_3)$                                                       \\
      \hline
      \multirow{7}{*}{full}
                    & 0              & 4                 & (0, 0, 0), (1, 1, 0), (2, 2, 0), (3, 3, 0)                                             \\
                    & 1              & 9                 & (0, 1, 1), (1, 0, 1), (1, 1, 1), (1, 2, 1), (2, 1, 1), (2, 2, 1), (2, 3, 1), (3, 2, 1) \\
                    &                &                   & (3, 3, 1)                                                                              \\
                    & 2              & 11                & (0, 2, 2), (1, 1, 2), (1, 2, 2), (1, 3, 2), (2, 0, 2), (2, 1, 2), (2, 2, 2), (2, 3, 2) \\
                    &                &                   & (3, 1, 2), (3, 2, 2), (3, 3, 2)                                                        \\
                    & 3              & 10                & (0, 3, 3), (1, 2, 3), (1, 3, 3), (2, 1, 3), (2, 2, 3), (2, 3, 3), (3, 0, 3), (3, 1, 3) \\
                    &                &                   & (3, 2, 3), (3, 3, 3)                                                                   \\
      \hline
      \multirow{4}{*}{lite}
                    & 0              & 4                 & (0, 0, 0), (1, 1, 0), (2, 2, 0), (3, 3, 0)                                             \\
                    & 1              & 6                 & (0, 1, 1), (1, 0, 1), (1, 2, 1), (2, 1, 1), (2, 3, 1), (3, 2, 1)                       \\
                    & 2              & 4                 & (0, 2, 2), (1, 3, 2), (2, 0, 2), (3, 1, 2)                                             \\
                    & 3              & 2                 & (0, 3, 3), (3, 0, 3)                                                                   \\
      \hline
      \multirow{4}{*}{level}
                    & 0              & 4                 & (0, 0, 0), (1, 1, 0), (2, 2, 0), (3, 3, 0)                                             \\
                    & 1              & 5                 & (0, 1, 1), (1, 0, 1), (1, 1, 1), (1, 2, 1), (2, 1, 1)                                  \\
                    & 2              & 5                 & (0, 2, 2), (1, 1, 2), (1, 2, 2), (2, 0, 2), (2, 1, 2)                                  \\
                    & 3              & 4                 & (0, 3, 3), (1, 2, 3), (2, 1, 3), (3, 0, 3)                                             \\
      \hline
   \end{tabular}
\end{table}

\begin{table}[h]
   \centering
   \caption{Tensor product paths for $L=4$}
   \label{tab:tp:path:L4}
   \setlength{\tabcolsep}{6pt}
   \small
   \begin{tabular}{cccl}
      \hline
      \textbf{Mode} & \textbf{$l_3$}                                                                         & \textbf{\# Paths} & \textbf{Paths} $(l_1, l_2, l_3)$                                                       \\
      \hline
      \multirow{10}{*}{full}
                    & 0                                                                                      & 5                 & (0, 0, 0), (1, 1, 0), (2, 2, 0), (3, 3, 0), (4, 4, 0)                                  \\
                    & 1                                                                                      & 12                & (0, 1, 1), (1, 0, 1), (1, 1, 1), (1, 2, 1), (2, 1, 1), (2, 2, 1), (2, 3, 1), (3, 2, 1) \\
                    &                                                                                        &                   & (3, 3, 1), (3, 4, 1), (4, 3, 1), (4, 4, 1)                                             \\
                    & 2                                                                                      & 16
                    & (0, 2, 2), (1, 1, 2), (1, 2, 2), (1, 3, 2), (2, 0, 2), (2, 1, 2), (2, 2, 2), (2, 3, 2)                                                                                                              \\
                    &                                                                                        &                   & (2, 4, 2), (3, 1, 2), (3, 2, 2), (3, 3, 2), (3, 4, 2), (4, 2, 2), (4, 3, 2), (4, 4, 2) \\

                    & 3                                                                                      & 17                & (0, 3, 3), (1, 2, 3), (1, 3, 3), (1, 4, 3), (2, 1, 3), (2, 2, 3), (2, 3, 3), (2, 4, 3) \\
                    &                                                                                        &                   & (3, 0, 3), (3, 1, 3), (3, 2, 3), (3, 3, 3), (3, 4, 3), (4, 1, 3), (4, 2, 3), (4, 3, 3) \\
                    &                                                                                        &                   & (4, 4, 3)                                                                              \\
                    & 4                                                                                      & 15                & (0, 4, 4), (1, 3, 4), (1, 4, 4), (2, 2, 4), (2, 3, 4), (2, 4, 4), (3, 1, 4), (3, 2, 4) \\
                    &                                                                                        &                   & (3, 3, 4), (3, 4, 4), (4, 0, 4), (4, 1, 4), (4, 2, 4), (4, 3, 4), (4, 4, 4)            \\
      \hline
      \multirow{5}{*}{lite}
                    & 0                                                                                      & 5                 & (0, 0, 0), (1, 1, 0), (2, 2, 0), (3, 3, 0), (4, 4, 0)                                  \\
                    & 1                                                                                      & 8                 & (0, 1, 1), (1, 0, 1), (1, 2, 1), (2, 1, 1), (2, 3, 1), (3, 2, 1), (3, 4, 1), (4, 3, 1) \\
                    & 2                                                                                      & 6                 & (0, 2, 2), (1, 3, 2), (2, 0, 2), (2, 4, 2), (3, 1, 2), (4, 2, 2)                       \\
                    & 3                                                                                      & 4                 & (0, 3, 3), (1, 4, 3), (3, 0, 3), (4, 1, 3)                                             \\
                    & 4                                                                                      & 2                 & (0, 4, 4), (4, 0, 4)                                                                   \\
      \hline
      \multirow{5}{*}{level}
                    & 0                                                                                      & 5                 & (0, 0, 0), (1, 1, 0), (2, 2, 0), (3, 3, 0), (4, 4, 0)                                  \\
                    & 1                                                                                      & 6                 & (0, 1, 1), (1, 0, 1), (1, 1, 1), (1, 2, 1), (2, 1, 1), (2, 2, 1)                       \\
                    & 2                                                                                      & 8                 & (0, 2, 2), (1, 1, 2), (1, 2, 2), (1, 3, 2), (2, 0, 2), (2, 1, 2), (2, 2, 2), (3, 1, 2) \\
                    & 3                                                                                      & 7                 & (0, 3, 3), (1, 2, 3), (1, 3, 3), (2, 1, 3), (2, 2, 3), (3, 0, 3), (3, 1, 3)            \\
                    & 4                                                                                      & 5                 & (0, 4, 4), (1, 3, 4), (2, 2, 4), (3, 1, 4), (4, 0, 4)                                  \\
      \hline
   \end{tabular}
\end{table}

\clearpage
\section{Algorithm to compute hyper moment}

\begin{algorithm}[H]  
   \caption{Iterative Calculation of Hyper Moment Tensors}
   \label{alg:basic:compute}
   \begin{algorithmic}[1]
      \Require Atomic moment $M_{ul}$; Maximum correlation degree $v_\text{max}$
      \State $\mathbf{H}_{ul}^1 \leftarrow \mathbf{M}_{ul}$
      \For{$v \gets 2$ to $v_\text{max}$}
      \State $\mathbf{H}^v_{ul_3,p} \gets \mathbf{H}^{v-1}_{ul_1} \hat\otimes \mathbf{M}_{ul_2}$
      \State $\mathbf{H}^v_{ul} \gets \sum_p W_p \mathbf{H}^v_{ul,p} $
      \EndFor
      \State $\mathbf{H}_{ul} \gets \sum_{v=1}^{v_\text{max}} W_l^v \mathbf{H}_{ul}^v$
   \end{algorithmic}
\end{algorithm}

\clearpage

\def\bibsection{\begingroup\renewcommand{\addcontentsline}[3]{}\section*{Supplementary References}\endgroup}
%